\documentclass[journal]{IEEEtran}

\usepackage[utf8]{inputenc}
\usepackage{textcomp}

\usepackage{amsmath,amssymb}
\usepackage{mathtools}
\usepackage{amsthm}
\usepackage{altvars}
\usepackage{multirow}
\usepackage{enumerate}
\usepackage{enumitem}
\usepackage{filecontents}
\usepackage[nolists,nomarkers,nofigures,notables]{endfloat}

\usepackage[hidelinks]{hyperref}
\usepackage[capitalise]{cleveref}
\crefname{equation}{}{}
\Crefname{equation}{Equation}{Equations}
\crefname{figure}{Fig.}{Figs.}

\usepackage{graphicx}
\usepackage{tikz}

\usepackage{xparse}
\usepackage{xspace}

\usepackage{todonotes}
\presetkeys{todonotes}{inline,prepend,caption={todo}}{}

\usepackage[normalem]{ulem}
\usepackage{bm}

\usepackage{algorithm}
\usepackage{algorithmic}
\usepackage{subfig}

\usepackage{lipsum}
\usepackage{amsfonts}
\usepackage{graphicx}
\usepackage{epstopdf}
\usepackage{algorithmic}
\ifpdf
  \DeclareGraphicsExtensions{.eps,.pdf,.png,.jpg}
\else
  \DeclareGraphicsExtensions{.eps}
\fi

\usepackage{bm}

\crefname{equation}{}{}
\Crefname{equation}{Equation}{Equations}
\crefname{figure}{Fig.}{Figs.}

\usepackage{pgfplots}


\usepgfplotslibrary{colorbrewer}

\pgfplotsset{compat=newest}

\newcommand{\new}[1] {{\color{black}#1}}
\newcommand{\newnew}[1]{{\color{black}#1}}

\newcommand{\newthree}[1] {{\color{black}#1}}
\newcommand{\newfour}[1] {{\color{black}#1}}

\newcommand{\comment}[1]{} 

\newcommand{\argmin}{\operatorname*{argmin}}

\newcommand{\tr}{\operatorname{tr}}
\newcommand{\I}{\mathbf{I}}

\newcommand{\R}{\mathbb{R}}
\newcommand{\C}{\mathbb{C}}

\newcommand{\W}{\bm{W}}

\newcommand{\V}{\bm{V}}

\newcommand{\X}{\bm{X}}
\newcommand{\A}{\bm{A}}
\newcommand{\B}{\bm{B}}
\newcommand{\CC}{\bm{C}}

\newcommand{\FF}{\bm{F}}
\newcommand{\F}{\bm{F}}

\newcommand{\PP}{\bm{P}}
\newcommand{\RR}{\bm{R}}
\newcommand{\Ss}{\bm{S}}
\newcommand{\U}{\bm{U}}

\newcommand{\tensor}[1]{\pmb{\mathcal{#1}}}
\newcommand{\slice}[1]{\overrightarrow{\pmb{\mathcal{#1}}}}
\newcommand{\tube}[1]{\overrightarrow{#1}}
\newcommand{\tP}{\pmb{*}}
\newcommand{\bcirc}{\texttt{bcirc}}

\newcommand{\unfold}{\texttt{unfold}}
\newcommand{\fold}{\texttt{fold}}
\newcommand{\fft}{\texttt{fft}}
\newcommand{\ifft}{\texttt{ifft}}
\newcommand{\conj}{\texttt{conj}}

\newcommand{\xmath}[1] {\ensuremath{#1}\xspace}

\newcommand{\vv} {\xmath{\bm{v}}}
\newcommand{\av} {\xmath{\bm{a}}}
\newcommand{\bv} {\xmath{\bm{b}}}
\newcommand{\cv} {\xmath{\bm{c}}}
\newcommand{\wv} {\xmath{\bm{w}}}

\newcommand{\fb} {\bar{\xmath{\bm{f}}}}

\newcommand{\ei} {\xmath{\bm{e}_i}}
\newcommand{\xb} {\xmath{\bm{\bar x}}}
\newcommand{\wb} {\xmath{\bm{\bar w}}}
\newcommand{\rb} {\xmath{\bm{\bar r}}}

\newcommand{\pb} {\xmath{\bm{\bar p}}}
\newcommand{\rhob} {\xmath{\bm{\bar \rho}}}
\newcommand{\gammab} {\xmath{\bm{\bar \gamma}}}
\newcommand{\sbt} {\xmath{\bm{\bar s}}_t}
\newcommand{\sbtk} {\xmath{\bm{\bar s}}_{t,k}}
\newcommand{\vb} {\xmath{\bm{\bar v}}}
\newcommand{\Ab} {\overline \A}
\newcommand{\Bb} {\overline \B}
\newcommand{\Cb} {\overline \CC}
\newcommand{\Ub} {\overline \U}
\newcommand{\Wb} {\overline \W}
\newcommand{\Rb} {\overline \RR}
\newcommand{\Sb} {\overline \Ss}
\newcommand{\Vb} {\overline \V}
\newcommand{\Xb} {\overline \X}

\newcommand{\FOmeg}{\bm{\mathcal{F}}_{\Omega_t}}

\newcommand{\Fkron}{\bm{\mathcal{F}}}

\newcommand{\tA} {\tensor{A}}
\newcommand{\tB} {\tensor{B}}
\newcommand{\tC} {\tensor{C}}
\newcommand{\tU} {\tensor{ U}}
\newcommand{\tW} {\tensor{ W}}

\newcommand{\tR} {\tensor{ R}}
\newcommand{\tS} {\tensor{ S}}
\newcommand{\tX} {\tensor{ X}}
\newcommand{\tV} {\tensor{ V}}

\newcommand{\tAb} {\tensor{\overline A}}
\newcommand{\tBb} {\tensor{\overline B}}
\newcommand{\tCb} {\tensor{\overline C}}
\newcommand{\tUb} {\tensor{\overline U}}
\newcommand{\tWb} {\tensor{\overline W}}
\newcommand{\tSb} {\tensor{\overline S}}
\newcommand{\tVb} {\tensor{\overline V}}
\newcommand{\tXb} {\tensor{\overline X}}

\NewDocumentCommand \cL { s s }
  {
    \IfBooleanTF{#1}
      { \IfBooleanTF{#2}
        { \bar{\mathcal{L}} }
        { \mathcal{L}_{\mathrm{c}} } }
      { \mathcal{L} }
  }
  
\newtheorem{theorem}{Theorem}[section]

\newtheorem{lemma}[theorem]{Lemma}
\newtheorem{remark}[theorem]{Remark}
\newtheorem{proposition}[theorem]{Proposition}
\newtheorem{definition}[theorem]{Definition}
  
\definecolor{mycolor}{RGB}{255,127,14}

\makeatletter
\def\thickhline{%
  \noalign{\ifnum0=`}\fi\hrule \@height \thickarrayrulewidth \futurelet
   \reserved@a\@xthickhline}
\def\@xthickhline{\ifx\reserved@a\thickhline
               \vskip\doublerulesep
               \vskip-\thickarrayrulewidth
             \fi
      \ifnum0=`{\fi}}
\makeatother

\newlength{\thickarrayrulewidth}
\newtheorem*{assumption*}{\assumptionnumber}
\providecommand{\assumptionnumber}{}
\makeatletter
\newenvironment{assumption}[2]
 {%
  \renewcommand{\assumptionnumber}{A#1$\mathcal{#2}$}%
  \begin{assumption*}%
  \protected@edef\@currentlabel{A#1$\mathcal{#2}$}%
 }
 {%
  \end{assumption*}
 }
\makeatother
\usepackage[switch]{lineno}

\usepackage{xr-hyper}
\makeatletter
\newcommand*{\addFileDependency}[1]{
  \typeout{(#1)}
  \@addtofilelist{#1}
  \IfFileExists{#1}{}{\typeout{No file #1.}}
}
\makeatother

\begin{document}

\clearpage \setcounter{page}{1}

\title{Grassmannian Optimization for Online Tensor Completion and Tracking with the t-SVD}

\author{%
  Kyle~Gilman,~\IEEEmembership{Student~Member,~IEEE,} Davoud~Ataee~Tarzanagh,~\IEEEmembership{Member,~IEEE}
  ~and~Laura~Balzano,~\IEEEmembership{Senior~Member,~IEEE}
  \thanks{%
    K. Gilman, D. Tarzanagh, \& L. Balzano are with the
      Department of Electrical Engineering and Computer Science,
      University of Michigan,
      Ann Arbor, MI, 48109 USA (e-mail: kgilman@umich.edu; tarzanaq@umich.edu; girasole@umich.edu).
  }%
  \thanks{
    The authors were supported in part by AFOSR YIP award FA9550-19-1-0026, NSF CAREER award CCF-1845076, and NSF BIGDATA award IIS-1838179.
  }%
  \thanks{
  © 2022 IEEE. Personal use of this material is permitted. Permission from IEEE must be obtained for all other uses, in
any current or future media, including reprinting/republishing this material for advertising or promotional purposes, creating
new collective works, for resale or redistribution to servers or lists, or reuse of any copyrighted component of this work in other
works.
  }
}
  
\maketitle
\begin{abstract}

We propose a new fast streaming algorithm for the tensor completion problem of imputing missing entries of a low-tubal-rank tensor using the tensor singular value decomposition (t-SVD) algebraic framework. We show the t-SVD is a specialization of the well-studied block-term decomposition for third-order tensors, and we present an algorithm under this model that can track changing free submodules from incomplete streaming {2-D} data. The proposed algorithm uses principles from incremental gradient descent on the Grassmann manifold of subspaces to solve the tensor completion problem with linear complexity and constant memory in the number of time samples. \newnew{We provide a local expected linear convergence result for our algorithm.} Our empirical results are competitive in accuracy but much faster in compute time than state-of-the-art tensor completion algorithms on real applications to recover temporal chemo-sensing and MRI data under limited sampling.

\end{abstract}



\section{Introduction}
\label{sec:intro}
Modern data are increasingly high-dimensional and multiway, increasing the storage and computational burden of signal processing algorithms. Many practical applications collect data over multiple modalities and can be approximated by a linear spectral mixture model, such as hyperspectral imaging (HSI), which captures dozens or even hundreds of images in narrow, adjacent spectral bands for each frame \cite{Wang}, or time-sequential HSI, i.e., hyperspectral video (HSV), with hundreds of spectral bands and megapixel spatial resolution which requires images to be recorded at the order of 10 G pixels per second. It is currently infeasible to process this type of high-rate data in real time applications \cite{Fletcher_Holmes_2005}. Similarly, chemo-sensing experiments record sensor readings from dozens of channels in hundreds of experiments over thousands of time series. Batch processing of large-scale tensor data quickly becomes computationally intractable, and even storing these tensors is problematic as the memory requirements grow rapidly with the number and size of the tensor modes. Additional challenges include large numbers of missing tensor entries, streaming multiway data that needs to be processed on the fly, and data that may evolve in time with model dynamics.

To address these concerns, there is extensive recent literature studying low-dimensional tensor decompositions and fast algorithms for computing them. These decompositions provide a low-memory model approximation to tensor data that can be used for compression and interpolation of missing entries. Several algebraic frameworks exist for the analysis and decomposition of tensors, each with their own notion of tensor rank. In this paper, we consider sampling and recovery of three-way tensors using the algebraic framework of the tensor singular value decomposition (t-SVD) \cite{Kilmer2011FactorizationSF,kilmer_braman_hao_hoover,martin_shafer_larue}. Three-way tensors are treated as linear operators over the space of oriented matrices and group rings of fibers under the tensor-product (t-product) multiplicative operator. Using this framework, one obtains an SVD-like factorization referred to as the tensor-SVD (t-SVD) with a defined notion of rank referred to as the tubal-rank. A key property of the t-SVD is the optimality of the truncated t-SVD for data approximation under the Frobenius norm measure \cite{Zhang_Aeron}. The t-SVD has found wide utility in computer vision \cite{Lu,online_2d_tensor_rpca}, image and signal processing \cite{Zhang_et_al,tctf_algorithm,tensor_comp_visual_data}, geophysics, HSI/HSV \cite{HSI_restoration,tv_hsi_denoise}, and other applications because of its ability to capture signal shifts and scaling due to the model's circulant algebra. However, most existing t-SVD based methods are batch methods that require all of the data to be stored in memory at computation time and/or require the computation of multiple SVDs. This is very time-consuming and inefficient for large-scale data. Current t-SVD algorithms also do not model dynamically changing data.

Despite much development of the t-SVD, little work has shown its connections to standard multilinear algebra models, which are more mathematically interpretable. In this paper, we show the t-SVD can be equivalently expressed in standard multilinear algebra as a certain block-term decomposition (BTD) problem. The BTD model is a generalization of both the CANDECOMP/PARAFAC decomposition (CPD) and Tucker tensor decompositions with important applications in linear spectral mixture models, decoupling multivariate polynomials, and audio signal separation \cite{qian_et_al_matrix-vector}. To the best of our knowledge, we are the first to show this equivalence. We show the t-SVD is an efficient factorization of each block in the BTD by utilizing a fixed unitary factor---the discrete Fourier transform matrix---in the third mode.

\sloppypar The impetus of this paper is to propose a fast, efficient algorithm for recovering low-tubal-rank tensor data from streaming, highly-incomplete multiway data with incremental gradient descent on the product manifold of low-rank matrices. Our methods are online by nature and can handle dynamically changing data, avoid computing SVDs, maintain orthonormality on the product of Grassmann manifolds, scale linearly in computation with the number of samples, and are highly parallelizable. We compare our method to batch t-SVD methods and online tensor decompositions. We show our method's ability to track dynamically time-varying low-rank free submodules in real data settings.

\subsection{Organization of this paper}

\begin{itemize}
\item Section \ref{sec:prelim} introduces our notation and the mathematical representations for the CP, Tucker, and BTD decompositions. Since our tensor factorization algorithm is based on the t-product \cite{Kilmer2011FactorizationSF}, we briefly cover the background for this decomposition and leave the details for the appendix. At a high level, the t-product is convolutional and so can be performed by a product in the Fourier domain. We also discuss the properties of the t-SVD as it relates to the block-term decomposition.

\item Section \ref{sec:related_work} details related work in tensor decompositions and completion.
\item Section \ref{sec:tsvd} proposes our tensor completion method, summarized in Algorithm \ref{alg:toucan}. \newnew{This section also provides a local convergence result for the proposed algorithm, showing that in a local region we achieve a linear convergence rate in expectation.} 
\item Section \ref{sec:experiments} gives experimental results for synthetic data, chemo-sensing experiments, and MRI completion.
\end{itemize}

\section{Preliminaries}
\label{sec:prelim}

\subsection{Notation}
\label{sec:prelim:subsec:notation}
We shall denote all scalar quantities as $s$, vectors as $\vv$, matrices as $\A$, and tensors as $\tensor{X}$. The $i^{th}$ lateral slice of a three-way tensor $\tensor{X}$ is a matrix and is denoted as $\slice{X}_i$; in MATLAB notation this object refers to $\tensor{X}(:,i,:)$. The (frontal) faces of a tensor $\tensor{X}$, denoted as $\X_i$, are $\tensor{X}(:,:,i)$. Any $1 \times 1 \times d_3$ tube along the third-dimension is denoted as $\tube{
\vv}$. The $n$-mode unfolding of a tensor $\tensor{X} \in \F^{d_1 \times \cdots \times d_N}$ into a $d_n \times \Pi_{i\neq n}^N d_i$ matrix is written as $\X_{(n)}$. 

We write the Kronecker product as $\otimes$, the Khatri-Rao product as $\odot$, and the outer product as $\circ$. The mode-$n$ product of a tensor $\tensor{X}$ with matrix $\CC$ is denoted as \newnew{$\tensor{X} \times_n \CC$ and its mode-$n$ matricization is defined as $(\tensor{X} \times_n \CC)_{(n)} = \CC \X_{(n)}$}. Refer to \cite{kolda_bader} for more on these products and their properties and identities. For the purposes of t-SVD and BTD models, we will often need to write $\tensor{X} \times_3 \CC$, for some tensor $\tensor{X}$ and matrix $\CC$, which we will denote as $\tXb$. The faces of $\tXb$ are then written as $\Xb_i$.

We denote the Frobenius norm as $\|\tA\|_F = \sqrt{\sum_{ijk} {|\tA_{ijk}|}^2}$. The complex conjugate of a quantity $\texttt{conj}(\cdot)$ takes the complex conjugate of each entry. The complex conjugate transpose of a matrix $\A$ is denoted as $\A'$ and the psuedo-inverse as $\A^{\dagger}$.

\subsection{Multilinear tensor decompositions}
A rank-$r$ CP decomposition is a sum of $r$ rank-1 outer products \cite{hitchcock_tensors_1927}. For a three-way tensor $\tX \in \F^{d_1 \times d_2 \times d_3}$ with scalar weights $\bmlambda = [\lambda_1 \cdots \lambda_r]' \in \R^{r}$ and factor matrices (assumed to be normalized to have unit column norms) $\A = [\av_1 \cdots \av_r] \in \F^{d_1 \times r}$, $\B = [\bv_1 \cdots \bv_r] \in \F^{d_2 \times r}$, and $\CC = [\cv_1 \cdots \cv_r] \in \F^{d_3 \times r}$ the decomposition is expressed as
\begin{equation}
    \text{(CP)} \qquad \tensor{X} \approx \sum_{i=1}^r \lambda_i \av_i \circ \bv_i \circ \cv_i := [\![ \bmlambda; \A, \B, \CC]\!].
\end{equation}

A multirank-$(m,n,p)$ Tucker decomposition \cite{tucker1966some} permits a different rank in each mode unfolding of the tensor and represents each \new{unfolding's columns} in the span of an orthonormal basis. The Tucker decomposition for orthonormal factor matrices $\A = [\av_1 \cdots \av_m] \in \F^{d_1 \times m}$, $\B = [\bv_1 \cdots \bv_n] \in \F^{d_2 \times n}$, $\CC = [\cv_1 \cdots \cv_p] \in \F^{d_3 \times p}$ and core tensor $\tensor{G} \in \F^{m \times n \times p}$ is 
\begin{equation}
    \text{(Tucker)} \qquad \tensor{X} \approx \tensor{G} \times_1 \A \times_2 \B \times_3 \CC := [\![\tensor{G}; \A, \B, \CC]\!]. 
    \label{eq:tucker}
\end{equation}

The core tensor is a smaller tensor whose entries show the level of interaction between the different components $\A, \B,$ and $\CC$. A Tucker tensor is decomposed as a core multiplied by the corresponding factor matrix along each mode \cite{kolda_bader}. Observe that the CPD is a Tucker tensor whose factors are non-orthogonal with a core tensor having all ones along the super-diagonal and zeros everywhere else.

The block-term decomposition (BTD) model \cite{de_lathauwer_lieven} is a useful generalization of both the CP and Tucker decompositions. The model expresses a third-order tensor $\tensor{X} \in \R^{d_1 \times d_2 \times d_3}$ as a sum of low-multirank tensors:
\begin{equation}
    \text{(BTD)} \qquad \tensor{X} \approx \sum_{k=1}^K [\![\tensor{G}_k; \A_k, \B_k, \CC_k ]\!],
    \label{eq:btd_model}
\end{equation}
where $\tensor{G}_k \in \R^{M_k \times N_k \times P_k}$ is each multirank-$(M_k,N_k,P_k)$ core tensor, and $\A_k \in \R^{d_1 \times M_k}$, $\B_k \in \R^{d_2 \times N_k}$, and $\CC_k \in \R^{d_3 \times P_k}$ for $k=1,\hdots,K$ are the factor matrices. From \cref{eq:btd_model}, it is easy to see the $K=1$ BTD with orthogonal factors specializes to the Tucker decomposition while a multirank-$(1,1,1)$ BTD is simply the CPD.

\subsection{t-SVD tensors}
\label{sec:tsvddefs}

\subsubsection{Discrete Fourier Transform}

Denote the normalized Discrete Fourier Transform (DFT) matrix for operation on a length-$n$ signal as the unitary matrix $\FF_n \in \C^{n \times n}$ and the DFT of some vector $\vv \in \R^n$ as $\vb = \FF_n \vv \in \C^n$. The DFT is computed in $\mathcal{O}(n\log n)$ time by the fast Fourier transform (FFT) as $\vb = \fft(\vv)$. Similarly, $\vv = \FF_n' \vb$ computes the inverse DFT (IDFT).

We denote $\tXb \in \C^{d_1 \times d_2 \times d_3}$ as the result of computing the DFT along the $3^{\text{rd}}$ dimension, i.e. performing the DFT on the tubes of $\tX$, or equivalently $\tXb := \tX \times_3 \FF_{d_3}$. Using the FFT (with indexing in MATLAB notation) we have $\tXb = \fft(\tX,[],3)$ and similarly by the inverse DFT, we have $\tX = \ifft(\tXb,[],3)$.

\subsubsection{Tensor-tensor product}
Define the block-diagonal matrix $\Xb \in \C^{d_1d_3 \times d_2d_3}$ to be the matrix with $d_3$ frontal faces of $\tXb$ along the diagonal, i.e. denote each frontal face of size $d_1 \times d_2$ as $\Xb_k$ and we have
\begin{equation}\label{eqn:bdiag}
    \Xb = \texttt{bdiag}(\tXb) = \begin{bmatrix} \Xb_1 & & \\ & \ddots & \\ & & \Xb_{d_3} \end{bmatrix}.
\end{equation}
\noindent We define the block-circulant matrix of the frontal faces of $\tX$ as $\bcirc(\tX)$, where 
\begin{equation*}
    \bcirc(\tX) = \begin{bmatrix} \X_1 & \X_{d_3} & \ldots & \X_2 \\ \X_2 & \X_1& \ldots & \X_3 \\ \vdots & \vdots & \ddots & \vdots \\ \X_{d_3} & \X_{d_3 - 1} & \ldots & \X_1 \end{bmatrix} \in \R^{d_1 d_3 \times d_2 d_3}.
\end{equation*}
\noindent From properties of block-circulant matrices, $\bcirc(\tX)$ can be block-diagonalized by the DFT:

\begin{equation}
    \Xb = (\FF_{d_3} \otimes \I_{d_1}) \cdot \bcirc(\tX) \cdot (\FF_{d_3}^{-1} \otimes \I_{d_2}),
    \label{block-diag}
\end{equation}
\noindent where $(\FF_{d_3}^{-1} \otimes \I_{d_2})$ is unitary \cite{Lu}.
For $\tX \in \R^{d_1 \times d_2 \times d_3}$ we define the fold and unfold operators \cite{Kilmer2011FactorizationSF}:
\begin{align*}
    \unfold(\tX) &= \begin{bmatrix} \X_1' & \X_2' & \cdots & \X_{d_3}' \end{bmatrix}', \nonumber \\
    \fold(\unfold(\tX)) &= \tX, \nonumber
\end{align*}
\noindent where the $\unfold(\cdot)$ operator maps $\tX$ to a matrix of size $d_1d_3 \times d_2$ and $\fold(\cdot)$ is its inverse operator. 

\begin{definition}{Tensor-product (t-product)}\cite{Kilmer2011FactorizationSF}: \textit{Let $\tA \in \R^{d_1 \times d_2 \times d_3}$ and $\tB \in \R^{d_2 \times l \times d_3}$. Then the t-product $\tA \tP \tB$ is defined to be a tensor of size $d_1 \times l \times d_3$,}
\label{def:tproduct}
\end{definition}
\begin{equation}
    \tA \tP \tB = \fold(\bcirc(\tA) \cdot \unfold(\tB)).
\end{equation}

The t-product can be understood from several perspectives. First, in the canonical domain, a three-way tensor of size $d_1 \times d_2 \times d_3$ can be thought of as an $d_1 \times d_2$ matrix whose entries are tubes lying in the third dimension. The t-product is then analogous to matrix-matrix multiplication but where circular convolution replaces scalar multiplication between the matrix elements. Second, the t-product is equivalent to matrix-matrix multiplication in the Fourier domain, or $\tC = \tA \tP \tB$ is equivalent to $\Cb = \Ab \, \Bb$ from \cref{block-diag}. This is shown as follows:
\begin{align}
\label{eq:t-product_Fourier}
&\unfold(\tC) = \bcirc(\tA) \cdot  \unfold(\tB) \nonumber \\
&= (\FF_{d_3}^{-1} \otimes \I_{d_1}) \cdot ((\FF_{d_3} \otimes \I_{d_1}) \bcirc(\tA) (\FF_{d_3}^{-1} \otimes \I_{d_2})) \\ &{\quad \hspace*{2em} \nonumber} \cdot ((\FF_{d_3} \otimes \I_{d_2}) \unfold(\tB)) \nonumber \\ 
&= (\FF_{d_3}^{-1} \otimes \I_{d_1}) \Ab \cdot \unfold(\tBb). \nonumber \end{align}
Therefore,
\begin{align*}
\unfold(\tCb) &= \Ab \cdot \unfold(\tBb)
\end{align*}
and for each front slice of $\Cb$, $\Cb_k = \Ab_k \Bb_k \quad \forall k=1,\hdots,d_3$.
Eq. \cref{eq:t-product_Fourier} and \cref{lemma:conj_symm} in the Appendix admit an efficient algorithm to compute the t-product using FFTs, as shown in \cref{alg:t-product} in the Appendix. Like matrix multiplication, the t-product is associative and linear \cite{Kilmer2011FactorizationSF}. In the case where $d_3 = 1$, it is easy to see that the t-product becomes regular matrix multiplication.

With the definition of this product between tensors, we can define analogous definitions of conjugate transpose ($\tensor{X}'$), the identity tensor ($\tensor{I}_{nnd} \in \R^{n \times n \times d}$), orthogonal tensors, and a type of diagonal tensor called the F-diagonal tensor. We leave these details for the reader in \cref{appendix:prelims}. Next we briefly discuss an SVD-like factorization of tensors under the t-product, and a definition of tubal-rank under the t-product and t-SVD.

\begin{theorem}[Tensor Singular Value Decomposition (t-SVD)] \cite{Kilmer2011FactorizationSF} \textit{Any tensor $\tA \in \R^{d_1 \times d_2 \times d_3}$ can be factorized as $\tA = \tU \tP \tS \tP \tV'$, where $\tU \in \R^{d_1 \times d_1 \times d_3}, \tV \in \R^{d_2 \times d_2 \times d_3}$ are orthogonal tensors, and $\tS \in \R^{d_1 \times d_2 \times d_3}$ is an F-diagonal tensor.}


\label{thm:tsvd}
\end{theorem}
We state \cref{thm:tsvd} without proof here and refer the reader to \cite{Lu} for a detailed proof. The t-SVD can be computed efficiently by \cref{alg:t-svd} in the appendix.

\begin{definition}[Tensor \new{multi-rank and} tubal-rank] \cite{tctf_algorithm}
\textit{For any $\tX \in \R^{d_1 \times d_2 \times d_3}$, \new{its multi-rank is a vector defined as $\bmr = (\textnormal{rank}(\Xb_1),\hdots,\textnormal{rank}(\Xb_{d_3})) \in \R^{d_3}$.} The tensor tubal-rank, $\textnormal{rank}_t(\tX)$, is defined as the number of nonzero singular tubes of $\tS$ from the t-SVD, i.e., \[\textnormal{rank}_t(\tX) = \#\{i : \tS(i,i,:) \neq \textbf{0}\}  \new{= \max \{r_1,\hdots,r_{d_3}\}}, \] \new{where $r_k = \textnormal{rank}(\Xb_k)$.}}
\end{definition}

\begin{definition}[Module over the commutative ring] \cite{online_2d_tensor_rpca} It can be shown the set of tubes $\C^{1 \times 1 \times d_3}$ equipped with the t-product forms a ring with unity $\R(\mathbb{G}_{d_3})$ \cite{tarzanagh_michailidis:18}. Define $\mathbb{M}^{d_1}_{d_3}$ to be a module, or the set of all 2-D lateral slices of size $d_1 \times 1 \times d_3$, over the ring of tubes. Since for any element $\slice{X} \in \mathbb{M}^{d_1}_{d_3}$ and coefficient tube $\tube{\vv} \in \R^{1 \times 1 \times d_3}$, the lateral slice $\slice{Y} = \slice{X} \tP \tube{\vv}$ is also an element of the module, so $\mathbb{M}^{d_1}_{d_3}$ is closed under tubal-scalar multiplication.
\end{definition}

\begin{definition}[Free submodule \new{(FSM)}]
$\mathbb{M}^{d_1}_{d_3}$ is called a free submodule of dimension $r < d_1$ over the commutative ring $\R(\mathbb{G}_{d_3})$ \cite{online_2d_tensor_rpca}, where one can construct a spanning basis of orthonormal lateral slices $\{\slice{U}_1,\slice{U}_2,\hdots,\slice{U}_{r}\}$ for which we we can uniquely represent any element $\slice{X} \in \mathbb{M}^{d_1}_{d_3}$ as a t-linear combination of the spanning basis with some tubal coefficients $\tube{\wv}_k$:
\begin{equation}
    \slice{X} = \sum_{k=1}^{r} \slice{U}_k \tP \tube{\wv}_k = \tU \tP \slice{W}.
    \label{eq:spanning_basis}
\end{equation}
Together, $\{\slice{U}_1,\slice{U}_2,\hdots,\slice{U}_{r}\}$ form the orthonormal tensor $\tU \in \R^{d_1 \times r \times d_3}$, and the arranged tubes $\tube{\wv}_k$ form the lateral slice $\slice{W} \in \R^{r \times 1 \times d_3}$.
\end{definition}

The definitions of free submodule over a ring generalize the notions of vector subspaces over a field of scalars and a spanning basis for a vector subspace, where the scalars of the field are the elements of the ring. Our algorithm uses the notions of free submodule to learn a spanning basis for the observed 2-D lateral slices of data in $\mathbb{M}^{d_1}_{d_3}$.

\newnew{

Before defining the manifolds and orthogonal groups used in our tensor problem, we first denote the complex orthogonal group, complex Stiefel manifold, and complex Grassmann manifold, respectively, from matrix linear algebra \cite{absil2009optimization,edelman1998geometry}:
\begin{align}\label{eqn:grassmannian_def}
\nonumber
\overline{\mathcal{O}}(r) &:= \{\Rb \in \C^{r \times r},~\Rb'\,\Rb = \Rb \,\Rb' = \I_r \}, \\
\nonumber
\overline{\mathcal{S}}(r,d_1) &:= \{\Ub'\Ub = \I: \Ub \in \C^{d_1 \times r}\}, \\
[\Ub] &:= \big\{\Ub \,\Rb: ~\Rb \in \overline{\mathcal{O}}(r)\big\} \in \overline{\mathcal{G}}(r,d_1),~\text{for}~\Ub \in \overline{\mathcal{S}}(r,d_1).
\end{align}
Next  we provide extensions of these definitions to third order tensors under the t-product.  
\begin{definition}[t-orthogonal group]
\label{def:tensor_ortho_group}
    Define $\mathcal{O}(r,r,d_3)$ as the t-orthogonal group of tubal rank-$r$:
    \begin{align}
         \mathcal{O}(r,r,d_3) &:= \big\{ \tR \in \mathbb{R}^{r \times r \times d_3}:~\tR' \tP \tR = \tR \tP \tR' = \tensor{I}_{rrd_3} \big\}.
    \end{align}
\end{definition}
\begin{definition}[t-Stiefel manifold]
The t-Stiefel manifold consisting of all tubal-rank-$r$ tensors with orthonormal lateral slices defined as 
\begin{equation}
\label{eq:t-product_stiefel}
\mathcal{S}(r, d_1, d_3) :=  \{\tU \in \mathbb{R}^{d_1 \times r \times d_3},~\tU' \tP \tU = \tensor{I}_{rrd_3} \}. 
\end{equation}
\end{definition}

We note that the t-Stiefel manifold is indeed a product of Stiefel manifolds in the Fourier domain, since each frontal slice of $\tUb = \tU \times_3 \FF_{d_3}$ is orthonormal and is a point on a Stiefel manifold, making $\tUb$ a point in the product space of Stiefel manifolds. We also note that $\tU = \tUb \times_3 \FF_{d_3}'$, where $\FF_{d_3}'$ is an invertible linear mapping of the frontal slices of $\tUb$. This together with the smoothness of $\tUb$ (as a product of smooth manifolds) implies that $\mathcal{S}(r, d_1, d_3)$ is also a smooth manifold; see, e.g.,~\cite[Lemma~1]{Zhang_Hou_Wang2020}.

\begin{definition}[t-Grassmann manifold]
\label{def:tensor_grassmannian}
Let $\sim_t$ denote an equivalence relation on the t-Stiefel manifold $\mathcal{S}(r, d_1,d_3)$ in the sense that for any $\tU_1, \tU_2 \in \mathcal{S}(r,d_1,d_3)$,  $\tU_1 \sim_t \tU_2 $ means that there exists a $\tR \in \mathcal{O}(r,r,d_3) $ such that $\tU_1 = \tU_2 \tP \tR$. The quotient space of $\mathcal{S}(r, d_1,d_3)$ under this equivalence relation, $\mathcal{S}(r, d_1,d_3)/\mathcal{O}(r,r,d_3)$, is called t-Grassmann manifold, i.e.,
\begin{align}
         \mathcal{G}(r,d_1,d_3) &:= \big\{  [\tU]:~\tU \in \mathcal{S}(r, d_1, d_3) \big\},
\end{align}
\end{definition}
where $[\tU]$ denotes the equivalence class under $\sim_t$:
\begin{align*}
    [\tU] := \big\{\tU \tP \tR &:~\tR \in \mathcal{O}(r,r,d_3) \big\},
\end{align*}
which is the $r$-dimensional free sub-module in $\mathbb{M}_{d_3}^{d_1}$ spanned under the t-product by the lateral slices of $\tU$.
}

\newthree{
\begin{proposition}
\label{propzer:well_defined_tensor_grassmannian}
$\mathcal{G}(r,d_1,d_3) $ is a smooth compact manifold of {
\color{black} dimension $d_3r(d_1-r)$}. 

\end{proposition}
}

\begin{proof}
Let
\begin{align}\label{eqn:gbar_def_article}
\begin{split}
&\overline{\mathcal{G}}(r,d_1,d_3) := \\
&\hspace{2mm}\left\{  [\tUb]:~ \tUb \in \C^{d_1 \times r \times d_3},~ \Ub_k\in \overline{\mathcal{S}}(r,d_1),~~ \forall k \in[d_3] \right\},
\end{split}
\end{align}
where 
\begin{align*}
[\tUb] := \left\{ \fold \left(\Ub_1 \Rb_1;\hdots; \Ub_{d_3}\Rb_{d_3}\right) ,\Rb_k \in \overline{\mathcal{O}}(r),~ \forall k \in[d_3] \right\},
\end{align*}
recalling $ \fold(\cdot) $  is defined in \cref{sec:prelim} and stacks the slices in its argument into a tensor.    
Then, we have
\begin{align*}
\nonumber
[\tUb] &=  \{\Ub_1 \Rb: ~\Rb \in \overline{\mathcal{O}}(r)\} \times \cdots \times  \{\Ub_{d_3} \Rb: ~\Rb \in \overline{\mathcal{O}}(r)\}\\
&= [\Ub_1] \times [\Ub_2] \times \cdots \times [\Ub_{d_3}].
\end{align*}
This implies that $[\tUb] \in \overline{\mathcal{G}}(r,d_1) \times \cdots \times \overline{\mathcal{G}}(r,d_1)$, where $\overline{\mathcal{G}}(r,d_1)$ is a smooth compact manifold of dimension \textcolor{black}{$r(d_1-r)$} \cite{edelman1998geometry}. Therefore, $\overline{\mathcal{G}}(r,d_1,d_3)$  is a smooth compact (product) manifold of dimension \textcolor{black}{$d_3 r(d_1-r)$}, and since the Fourier transform is invertible, using the properties of the t-product in \cref{def:tproduct}, for any ${\tU \in \mathcal{S}(r, d_1, d_3)}$, we have
${[\tU \times_3 \FF_{d_3}] = [\tUb]}$ and ${[\tUb \times_3 \FF_{d_3}'] = [\tU]}$. This implies that the t-Grassmannian $\mathcal{G}(r, d_1, d_3)$ from Definition \ref{def:tensor_grassmannian} is indeed a smooth and compact manifold. 

\end{proof}

\subsection{t-SVD and BTD Equivalence}

\begin{figure}
    \centering
    \includegraphics[width=0.45\textwidth]{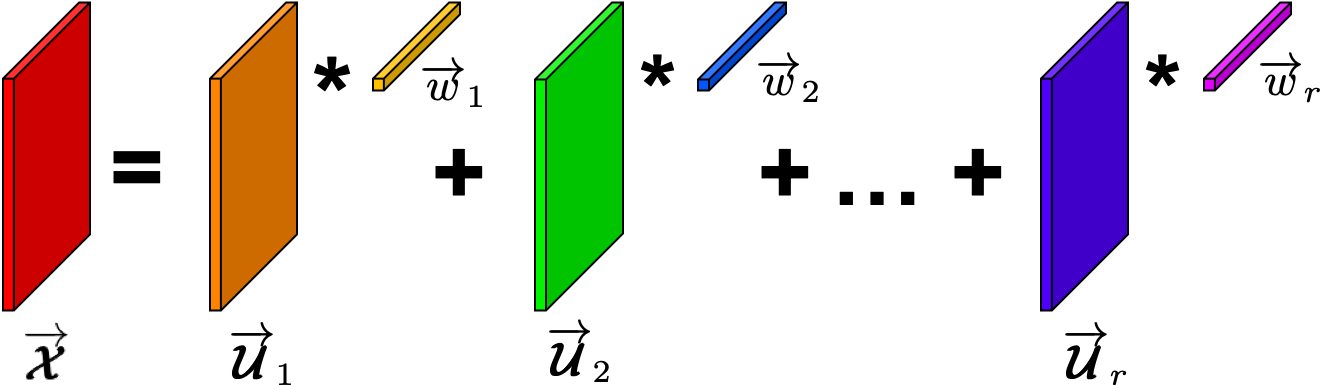}
    \caption{An element of a free module generated by t-linear combination of spanning basis and coefficients.}
    \label{fig:free_module_span_basis}
\end{figure}

\new{
\begin{definition}[$\{(r_k,r_k,1)\}_{k=1}^K$-multi-rank BTD]
    For multi-rank vector $\bmr = [r_1 \cdots r_K]' \in \mathbb{N}^{K}_+$, define the BTD tensor with the following decomposition for factors $\A_k \in \R^{d_1 \times r_k}$, $\B_k \in \R^{d_2 \times r_k}$, $\bmc_k \in \R^{d_3}$:
    \begin{align} \label{eq:multirank_btd}
        \tX = \sum_{k=1}^K \A_k \B_k' \circ \bmc_k.
    \end{align}
\end{definition}
}

\new{
\begin{proposition}

Let $\FF_{d_3}^{-1} = [\fb_1 \cdots \fb_{d_3}]$, where $\FF_{d_3}^{-1}$ denotes the inverse-DFT matrix. For any tensor decomposition $\tX = \tU \tP \tW$ with factor tensors  $\tU \in \R^{d_1 \times r \times d_3}$, $\tW \in \R^{r \times d_2 \times d_3}$, multi-rank $(r_1,\hdots,r_{d_3})$, and tubal rank $r= \max_{i} \{r_i\}$, we have
\begin{equation}\label{eq:batch_btd_model}
   \tU \tP \tW = \sum_{k=1}^{d_3} \Ub_k \Wb_k' \circ \fb_k.
\end{equation}
Here, $\Ub_k \in \C^{d_1 \times \new{r_k}}$, $\Ub_k' \Ub_k = \I_\new{{r_k}}$, $\Wb'_{k} \in \C^{\new{r_k} \times d_2} \quad \forall k=1,\hdots,d_3$, and $\Ub_k$ and $\Wb_{k}'$ are the \new{rank-$r_k$} faces of $\tUb = \tU \times_3 \FF_{d_3}$ and $\tWb = \tW \times_3 \FF_{d_3}$ respectively. 


\label{thm:tsvd_btd_equivalency}
\end{proposition}

\begin{proof}
    Let $\FF_{d_3}^{-1}$ be the $d_3 \times d_3$ IDFT matrix. The identity in \cref{eq:batch_btd_model} is clear after writing the definition of the t-product:
    \begin{align*}
        &\unfold(\tensor{U} \tP \tensor{W}) = (\FF_{d_3}^{-1} \otimes \I_{d_1})\cdot \\
        &(\FF_{d_3} \otimes \I_{d_1}) \bcirc(\tensor{U}) (\FF_{d_3}^{-1} \otimes \I_{d_2})\cdot(\FF_{d_3} \otimes \I_{d_2}) \unfold(\tensor{W})\\
        &= (\FF_{d_3}^{-1} \otimes \I_{d_1})\cdot \Ub \cdot \unfold(\Wb) \\
        &= \unfold\left(\sum_{k=1}^{d_3} (\Ub_k \Wb_k') \circ \fb_k \right).
    \end{align*}
    Note that $\tUb = \tU \times_3 \FF_{d_3}$ and $\tWb = \tW \times_3 \FF_{d_3}$ respectively; these mode-products apply the Fourier transform to the third-mode fibers.
\end{proof}

}


\new{In particular, Equation~\eqref{eq:batch_btd_model} links tubal and BTD decompositions and shows how a tensor factorization with multi-rank $(r_1,\hdots,r_{d_3})$ can equivalently be represented as a BTD factorization with multi-rank  $\{(r_k,r_k,1)\}_{k=1}^{d_3}$.} The equivalence reveals the t-SVD as a specialization of the BTD model with the third-mode fixed as the columns of the inverse DFT matrix. Each term in the t-SVD/BTD is itself a \new{($r_k,r_k,1$)-multi-rank} tensor with the identity core. In the linear spectral mixture model, each $\Ub_k \Wb_k'$ is the \new{rank-$r_k$} spectral map corresponding to a frequency component $\fb_k$. For a tubal-rank-1 t-SVD decomposition, \cref{eq:batch_btd_model} is a rank-$d_3$ CPD $[\![ \Ub; \Wb; \CC]\!]$, where $\CC = \FF_{d_3}^{-1}$ is the IDFT matrix.

\new{Choosing all $d_3$ of the multi-ranks would be especially challenging. Since the tubal-rank $r$ is much smaller than either of the tensor dimensions, we will slightly overparameterize the problem to a \textit{$(r,r,1)$-tubal BTD}:
\begin{definition}[$(r,r,1)$-tubal BTD]
Tensor $\tX$ has a $(r,r,1)$-tubal BTD if $\tX$ has the decomposition in \cref{eq:batch_btd_model} for 
$$  \tensor{X} = \sum_{k=1}^{d_3} \Ub_k \Wb_k' \circ \fb_k, $$
where 
$\Ub_k \in \C^{d_1 \times r}$, $\Ub_k' \Ub_k = \I_r$, $\Wb'_{k} \in \C^{\new{r} \times d_2} \quad \forall k=1,\hdots,d_3.$
\end{definition}

}

In the strict sense, the t-SVD is not a true tensor decomposition since it lacks the trilinearity in the third mode. Rather, it is a collection of matrix factorizations that describe a tensor structure. In the light of its relationship to the BTD and linear spectral mixture models, the t-SVD, along with its variants using the DCT or other invertible linear transforms \cite{KERNFELD2015545}, becomes appropriate for applications like HSI and video compression when the tensor faces are shifted and scaled versions of one another; i.e. the model describes spectral correlations by the embedded circular convolution \new{\cite{zheng2020}}. In the BTD form, the t-SVD decomposes the data into the frequency makeup of each pixel.

It is well-established the t-SVD is powerful in capturing the ubiquitous spatial-shifting and scaling correlations in real-world multiway data. We show tensors whose third modes exhibit these correlations lie in the t-linear span of the same free submodule, \new{and formalize this notion in \cref{prop:btd_tsvd_circular} using the multilinear algebra interpretation of the t-SVD.}
\begin{proposition}
\label{prop:btd_tsvd_circular}
    An \new{$r$-dimensional} free submodule \new{spanned by} $\tU$ over the t-product is closed under circular-shifting and scaling.
    \begin{proof}
        Let $\tX_i = \tU \tP \slice{W}_i = \sum_{k=1}^{d_3} \Ub_k \wb_{i,k} \circ \fb_k$\new{, where $\wb_{i,k} \in \C^{r}$ are the frontal faces of $\slice{W}_t \times_3 \FF_{d_3}$.}
        
        Let $\tX_{i,\text{shift}}:= \alpha_i \cdot \texttt{circshift}(\tX_i,s_i,\texttt{dims}=3)$ for some real numbers $\alpha_i$ and integers $s_i$ that scale and circularly shift the faces of each $\tX_i$. Then for $n$ linear combinations of slices,
        \begin{align}
            \sum_{i=1}^n \tX_{i,\text{shift}} &= \sum_{i=1}^n \sum_{k=1}^{d_3} \alpha_i e^{\frac{-j2\pi s_i k}{d_3}} \cdot \Ub_k \wb_{i,k} \circ \fb_k \nonumber \\
            &=  \sum_{k=1}^{d_3} \Ub_k \left(\sum_{i=1}^n \alpha_i e^{\frac{-j2\pi s_i k}{d_3}}  \wb_{i,k}\right) \circ \fb_k. \nonumber
        \end{align}
        Thus, $\sum_{i=1}^n \tX_{i,\text{shift}}$ shares the same $\tU$ in its t-SVD as each $\tX_i$.
    \end{proof}
\end{proposition}

\section{Related Work}
\label{sec:related_work}
It is well known that low-rank decompositions of highly undersampled matrix data, with certain assumptions of incoherent left and right singular vectors from the SVD and random sampling patterns, can be exploited to recover missing data by solving a convex optimization program \cite{candes_recht}. This setting treats matrix data (a 2-way tensor) as a linear operator over a vector space and defines the rank of the matrix via its minimal decomposition into a sum of rank-1 matrices \cite{Zhang_Aeron}. However, multiway data often contains correlations or interactions between modes of the tensor that would be destroyed if the tensor is flattened into a matrix \cite{tensors_book}. More sophisticated algebraic techniques are required to analyze these special structures. 

\subsection{CANDECOMP/PARAFAC decomposition}
One of the most widely used tensor decompositions is the CPD factorization, which finds a sum of rank-1 outer products that best compose the tensor, where the minimal number of such factors required is referred to as the CP rank. CP is powerful for imputing missing tensor data and also recovering latent factors that describe the tensor along each mode \cite{kolda_bader}. CP methods often use alternating least squares to update the factor matrices in a nonconvex optimization problem. Several varieties of CP algorithms exist for batch tensor completion \cite{cp_wopt,tenals,kolda_hong}. However there are known computational and ill-posedness issues with the CP model, the foremost issue being that it is NP-hard to compute the CP rank of a tensor or the best low-rank CP approximation of a tensor in the Frobenius norm sense \cite{hillar_lim_2013}. Furthermore, the alternating least squares algorithm is prone to getting stuck in local minima, so it may be sensitive to initialization or may require a special initialization step. CP models may also not be expressive enough to represent certain physical systems with block term decompositions.

Newer work in tensor completion has seen the development of several streaming CP tensor completion methods. A prominent streaming version of CP tensor completion was proposed by Mardani et al. \cite{mardani} using stochastic gradient descent. Kasai \cite{olstec} proposed another streaming CP tensor completion algorithm with a second-order stochastic gradient descent procedure based on the CP decomposition exploiting recursive least squares for faster convergence than the SGD method in Mardani et al., but at a higher computational cost. The main disadvantage to these streaming CP methods is that they require several hyperparameters that may be difficult to tune or know beforehand. These include a forgetting factor and the regularization parameters that penalize the Frobenius norm of the factor matrices \cite{tensor_completion_big_data_analytics}. While the forgetting factor must be hand-tuned, it does allow for the benefit of varying the algorithm's tracking ability from online mode to purely batch mode. Setting the appropriate CP rank of the model may also be challenging. Other streaming CP algorithms include \cite{maehara, samBaTen, ma_yang_wang, Vandecappelle, nion_Sidiropoulos}.

\subsection{Tucker decomposition}
\label{sec:rel:subsec:tucker}
Another approach is to use the Tucker tensor decomposition in \cref{eq:tucker} and Tucker multilinear rank (or multirank) and its convex relaxation. The multirank formulation allows each tensor mode to be expressed in \new{a subspace of different dimension}. 
\newnew{Tucker decompositions are typically computed using the Higher Order SVD (HOSVD) \cite{de2000multilinear}.}
However, Tucker-based convex relaxation is not a tight relaxation of the Tucker rank and cannot give optimal recovery for tensor completion \cite{tarzanagh_michailidis:18,Zhang_Aeron}.

The work in \cite{sun_et_ak_tucker_streaming} proposes using randomized linear algebra in the fully-observed data setting to sketch the Tucker decomposition, which naturally permits their algorithm to handle streaming data. The authors in \cite{Nimishakavi_tucker_streaming} propose a multi-aspect streaming Tucker-tensor algorithm for completing missing entries where one or more modes of the tensor grows in dimension length with time.

The online algorithm for tensor completion in \cite{stc} can also be thought of as an incremental Tucker algorithm with identity core tensor, \newnew{which is the tensor of ones along the super-diagonal and zero elsewhere}. Similar to our method, their algorithm tracks a low-\new{dimensional} subspace on the Grassmannian in each mode of the tensor using geodesic steps like the GROUSE algorithm \cite{grouse}.

\subsection{Block-term decomposition}

De Lathauwer et al. \cite{de_lathauwer_lieven} explores a special class of third-order tensor decompositions called the block-term decomposition (BTD) model and its theoretical properties, including the specialization to the multirank-$(r_k,r_k,1)$ model which represents a sum of matrix-vector outer
products. In many applications, it is natural to represent data in the BTD model in modes of space-space-frequency \cite{spectrum_cartography}. The works in \cite{qian_et_al_matrix-vector,spectrum_cartography} explicitly link the $(r_k,r_k,1)$-BTD to applications with strong physical interpretation like linear spectral mixture models and spectrum cartography that could not be well-represented by CP or Tucker tensors. BTD permits a richer, more expressive representation of data with more than one tensor component, like in the Tucker model, or without restriction to rank-1 components like CP \cite{qian_et_al_matrix-vector}.

The work in \cite{qian_et_al_matrix-vector} proposes a block-coordinate descent batch algorithm to compute the decomposition under full sampling. The authors of \cite{spectrum_cartography} propose algorithms for the case where tensor entries are missing in various patterns, and they prove the uniqueness and completion guarantees of the $(r_k,r_k,1)$ BTD factors under mild conditions.

\subsection{t-SVD}
The t-SVD, a factorization originally posed by \new{Kilmer} et al. \cite{kilmer_braman_hao_hoover}, enjoys many similar properties as matrix factorization problems, is solved by the SVD, and gives optimal recovery results under the Frobenius norm whenever the tensor data reveals a low-tubal rank structure \cite{Zhang_Aeron,Lu}. In many applications, for example time series or other ordered data, the corresponding tensor has a distinguishing orientation that exhibits a low tubal-rank structure \cite{tarzanagh_michailidis:18}. Several works have proposed t-SVD factorization algorithms for tensors with missing entries.  Zhang and Aeron \cite{Zhang_Aeron} solve the exact tensor completion problem under the t-SVD algebra in a batch way using the tensor nuclear norm, a convex relaxation of tensor tubal-rank. The algorithm involves solving a convex program on each frontal slice of the tensor in the Fourier domain, which provably recovers the missing tensor entries given certain incoherence conditions. Zhou et al. \cite{tctf_algorithm} propose a different algorithm using a tensor factorization model under the t-product for rapid, efficient optimization, and Tarzanagh and Michailidis \cite{tarzanagh_michailidis:18} employ randomized linear algebra to compute fast sketches of factorizations under the t-product. Each of these algorithms can only complete batch tensor data and cannot handle streaming multiway data.

Little work has been done to extend online matrix completion methods to the case of multiway tensor data using the t-SVD framework, apart from the work in \cite{online_2d_tensor_rpca} which proposed an online tensor robust principal component analysis algorithm. However, this method cannot predict missing tensor values and does not utilize orthonormal factorization. The work in \cite{pothier2015algorithm} proposed an online tensor completion algorithm using the tensor nuclear norm for low-tubal-rank tensors, but it must compute multiple SVDs for each update.

The algorithm proposed \new{in this paper} differs from all of these methods in that it can operate incrementally over a tensor in batch mode or stream in online mode, even with dynamically changing data. \new{Our proposed algorithm} TOUCAN seeks the optimal low-rank approximation of a tensor in the Frobenius norm sense under the t-SVD and the BTD models when the data reveals a low tubal-rank structure, and is empirically robust to initialization. TOUCAN requires only a tolerance threshold and the model rank---which can be more easily determined empirically by inspecting the tubal singular value decomposition of the t-SVD of small batches of data or over the entire batch if feasible. This paper builds off the work in \cite{gilman2020}, giving a full derivation of the algorithm, exploring connections to the BTD model, adding new algorithms and theory, and including new and more extensive experiments. 

\section{Proposed Method}
\label{sec:tsvd}
\subsection{Model}
\label{sec:methods:model}
In the t-SVD framework, using \cref{thm:tsvd_btd_equivalency} we model the three-way tensor data $\tensor{X} \in \mathbb{R}^{d_1 \times d_2 \times d_3}$ as
\begin{equation}
\label{eq:batch_tsvd_model}
    \tensor{X} \approx \sum_{k=1}^{d_3} \Ub_k \Wb_k' \circ \fb_k + \tensor{N} = \tensor{U} \tP \tensor{W} + \tensor{N},
\end{equation}
where $\tensor{N}_{ijk} \sim \mathcal{N}(0,\sigma^2)$ represents white-Gaussian noise, and $\tU \in R^{d_1 \times r \times d_3}$ is an orthonormal tensor under the tensor-product, and $\Ub_k \in \R^{d_1 \times r}, \Ub_k'\Ub_k = \I_r \quad \forall k=1,\hdots,d_3$.


Given $d_2$ 2-D data samples $\slice{X}_1,\hdots,\slice{X}_{d_2}$ of size $d_1 \times d_3$, we arrange them as lateral slices to make a three-way tensor $\tensor{X}$ of size $d_1 \times d_2 \times d_3$ \cite{online_2d_tensor_rpca}. In most circumstances, the t-SVD method would be used to compute \new{$\tU$ and $\tW$}  \cite{kilmer_braman_hao_hoover}. For the purposes of this work, we consider the case of three-way tensor data where each lateral slice arrives sequentially in time and may contain missing entries, i.e. at every time $t$, we observe an incomplete lateral slice $\slice{X}_t \in \mathbb{M}^{d_1}_{{d_3}}$ on the indices $\Omega_t \subset \{1,\hdots,d_1\} \times \{1,\hdots,d_3\}$. Like the work in \cite{online_2d_tensor_rpca}, we wish to compute the spanning low-dimensional free submodule of this multiway streaming data in an online way without storing the full tensor in memory or computing the t-SVD -- both which may be prohibitive in large data settings.

We can learn the spanning free submodule using stochastic gradient techniques, similar to what the GROUSE algorithm \new{\cite{grouse}} does for matrices with streaming columns. We aim to track a $r$-dimensional free submodule of $\mathbb{M}^{d_1}_{d_3}$ that may evolve over time. Let $\tensor{U} \in \mathbb{R}^{d_1 \times r \times d_3}$ be an orthonormal tensor whose $r$ lateral slices span \new{the} free submodule of $\mathbb{M}^{d_1}_{d_3}$.

\label{sec:method}

\subsection{Deriving the objective function}
\label{subsec:derive_obj_fxn}

\new{We begin by writing the problem we wish to solve as a tubal-rank-$r$ problem in t-SVD notation, and then we will express it as a $(r,r,1)$-tubal BTD}. In the scenario where the underlying free submodule does not change over time, a natural optimization problem with squared $\ell_2$ error loss is given as 
\newthree{
\begin{equation}
\begin{aligned}
    \label{eq:toucan_objective_incremental2}
   & \min_{[\tensor{U}] \in {\mathcal{G}(r,d_1,d_3)}}   \frac{1}{T} \sum_{t=1}^T \min_{\slice{W}_t \in \R^{r \times 1 \times d_3}} \frac{1}{2}\left\|\mathcal{A}_{\Omega_t}(\slice{X}_t - \tensor{U} \tP \slice{W}_t) \right\|^2_F. 
\end{aligned}
\end{equation}
}
Here, $\mathcal{A}_{\Omega_t}(\cdot)$ is the linear operator that extracts the observed samples in the set $\Omega_t$ from each lateral slice in $\tensor{X} = [\slice{X}_1 , \hdots, \slice{X}_{d_2}]$, and $\mathcal{G}(r,d_1,d_3)$ denotes the t-Grassmannian from \cref{def:tensor_grassmannian}. We let $\clL(\tU) := \frac{1}{T} \sum_{t=1}^T \cL_t(\tU)$ where 
\newthree{
\begin{subequations}
 \begin{align}
 \cL_t(\tU) := \frac{1}{2}\left\|\mathcal{A}_{\Omega_t}\left(\slice{X}_t - \tensor{U} \tP \slice{W}_t(\tensor{U}) \right)\right\|^2_F,~~\textnormal{and}~~ \label{eq:bilevel1}\\
 \slice{W}_t(\tensor{U}):=\argmin_{\slice{W}_t \in \R^{r \times 1 \times d_3}}~~\frac{1}{2}\left\|\mathcal{A}_{\Omega_t}\left(\slice{X}_t - \tensor{U} \tP \slice{W}_t \right)\right\|^2_F~\label{eq:bilevel2}. 
 \end{align}
\end{subequations}
}

Since we have concatenated the time slices on the second dimension, let $d_2 = T$. We see it is possible to solve this problem incrementally, as described in \cite{bertsekas}, in terms of the orthonormal free-submodule $\tU$ and the \newthree{optimal weights $\slice{W}_t(\tU) \in \mathbb{R}^{r \times 1 \times d_3}$} for all  $t=1,\hdots,T$. We solve the nested optimization problem in \cref{eq:toucan_objective_incremental2} for each slice $\slice{X}_t$ with stochastic gradient descent. From the results in \cref{thm:tsvd_btd_equivalency} and \cref{prop:well_defined_tensor_grassmannian}, we express each $\clL_t(\tU)$ in each slice at time $t$ as a $(r,r,1)$-tubal BTD. Let $\overline{\clL}_t$ denote $\cL_t$ in terms of the Fourier variables, and recall $\Ub$ denotes the block-diagonal matrix representation of $\tUb = \tU \times_3 \FF_{d_3}$, where $\Ub_k$ is the $k^{th}$ block on its diagonal of sizes $d_1 \times r$. Let $\Ub_k$ and \newthree{$\wb_{t,k}(\Ub) \in \C^{r}$} for all  $k\in[d_3]$ be the frontal faces of the tensors $\tUb$ and \newthree{the optimal $\slice{W}_t(\tU) \times_3 \FF_{d_3}$}, respectively. Then denoting \newthree{$\wb_t (\Ub)  = [\wb_{t,1}'(\Ub)   \cdots  \wb_{t,d_3}'(\Ub)]'$}, we can write $\overline{\clL}_t$ as
\begin{align*}
   \label{eq:arb_missing:btd_stochastic_objective2}
   &\overline{\cL}_t(\Ub) =  \frac{1}{2}\left\|\PP_{\Omega_t} \texttt{vec}\big(\slice{X}_t - \sum_{k=1}^{d_3} (\Ub_k \wb_{t,k}(\Ub) ) \circ \fb_r \big)\right\|_2^2 \notag \\
   &\new{=  \frac{1}{2} \big\|\PP_{\Omega_t}\texttt{vec}(\new{\Delta_{\Omega_t}(\slice{X}_t)})}\\ &{ \nonumber}
  \new{- \PP_{\Omega_t} \underbrace{\begin{bmatrix} (\fb_1 \otimes \I_{d_1}) & \hdots & (\fb_{d_3} \otimes \I_{d_1}) \end{bmatrix}}_{(\FF_{d_3}^{-1} \otimes \I_{d_1})} \begin{bmatrix}  \Ub_1 \wb_{t,1} (\Ub)  \\ \vdots \\ \Ub_{d_3} \wb_{t,d_3}(\Ub)\end{bmatrix}\big\|_F^2 }\\
   &= \frac{1}{2} \Big\|\FOmeg\big(\begin{bmatrix} \xb_{\Omega_t,1} \\ \vdots \\ \xb_{\Omega_t,d_3} \end{bmatrix} - \begin{bmatrix} \Ub_1 & & 0 \\& \ddots &  \\ 0 & & \Ub_{d_3} \end{bmatrix} \begin{bmatrix} \wb_{t,1}(\Ub) \\ \vdots \\ \wb_{t,d_3}(\Ub) \end{bmatrix}\big)\Big\|_2^2. \notag 
\end{align*}

Above, each $\xb_{\Omega_t,k} \in \C^{d_1}$ denotes the $k^{th}$ frontal face of $\tXb_{\Omega_t} = \new{\Delta_{\Omega_t}(\slice{X}_t)} \times_3 \FF_{d_3}$, where $\Delta_{\Omega_t}(\cdot)$ imputes zeros on the missing coordinates. $\PP_{\Omega_t}$ is a subsampled identity matrix of size $|\Omega_t| \times d_1d_3$, and $\FOmeg := \PP_{\Omega_t} (\FF_{d_3}' \otimes \I_{d_1}) \in \C^{|\Omega_t| \times d_1d_3}$, which in the t-SVD framework is the subsampled inverse Fourier transform. 
The derivation of this relation using the t-SVD algebra is also shown in \cref{appendix:sec:tsvd_toucan}.

Let us denote $\xb_t := \texttt{vec}(\tXb_{\Omega_t}) \in \C^{d_1d_3}$. Using the result above, the objective \cref{eq:toucan_objective_incremental2} in t-product form then has the equivalent nested optimization problem in the Fourier domain:
\newthree{
\begin{equation}
\begin{aligned}
   \label{eq:arb_missing:btd_stochastic_objective2}
    &\min_{[\Ub_k] \in \overline{\mathcal{G}}(r,d_1)~\forall k\in[d_3]} \frac{1}{T} \sum_{t=1}^T \overline{\cL}_t(\Ub),\\ &\textnormal{where}~~~ \overline{\cL}_t(\Ub) := \frac{1}{2}\|\FOmeg (\xb_t - \Ub \wb_t(\Ub))\|^2_2\\
    &\text{and}~~\wb_t(\Ub) =  \argmin_{\wb_t \in \mathbb{C}^{d_3r}} \frac{1}{2} \|\FOmeg (\xb_t - \Ub\wb_t)\|^2_2.
\end{aligned}
\end{equation}
}
The following proposition characterizes the smoothness of $\clL(\tU)$. 
\textcolor{black}{
\begin{proposition}
\label{prop:well_defined_tensor_grassmannian}
Suppose $|\Omega_t|$ is sufficiently large  such that $(\FOmeg \Ub)'(\FOmeg \Ub)$ remains full rank. Then,
\begin{enumerate}[label={\textnormal{(P\arabic*)}}]
 \item  \label{itm:prop:one} the inner least-squares problem \cref{eq:bilevel2} has a unique solution. 
    \item  \label{itm:prop:two}   $\clL$ is a well-defined smooth function over  $\mathcal{G}(r,d_1,d_3)$ and it admits a global optimizer. 
\end{enumerate}
\end{proposition}
We leave the proof of \cref{prop:well_defined_tensor_grassmannian} to \cref{appendix:proof:prop:well_defined_tensor_grassmannian}. At a high level, the proof shows that the (outer) optimization problem constrained to $[\tU] \in \mathcal{G}(r,d_1,d_3)$ is a well-defined problem on a product manifold of Grassmannians in the Fourier domain. Note that our assumption on $|\Omega_t|$ is identical to \cite{grouse, balzano_wright2015}.
}

The problem in \cref{eq:arb_missing:btd_stochastic_objective2} is nonconvex from the coupling of $\Ub$ and $\wb_{\Ub}$ and the orthonormality constraints $\Ub_k'\Ub_k = \I_r$ for all $k \in [d_3]$. We will minimize a problem over a product of $d_3$ Grassmannians $\mathcal{G}_1(r,d_1) \times \hdots \times \mathcal{G}_{d_3}(r,d_1)$ represented by $\Ub$. For a single data observation, we first compute the unique minimizer $\wb_t(\Ub)$ (equivalently $\slice{W}_t(\tU)$) to the inner least-squares problem, and then take a stochastic gradient descent step in the negative gradient direction of $\overline{\cL}_t(\Ub)$ with respect to each block $\Ub_k$ on the diagonal of $\Ub$ for estimating $\tUb$ (equivalently $\tU$).

\subsection{Updating \texorpdfstring{$\tensor{U}$}{U}}
\label{subsec:U_update}

To update our estimate of the free submodule $\tU$, we perform a gradient descent step on each Grassmannian in the Fourier domain. We compute the gradient of the objective function $\overline{\cL}_t$ with respect to each $\Ub_k$ and then follow this gradient along a short geodesic curve on the Grassmannian \cite{grouse}. \textcolor{black}{
Substituting \newfour{the expression for} the unique closed-form solution $\wb_t(\Ub)$ into the objective, where $\wb_t(\Ub) =  \argmin_{\wb_t \in \mathbb{C}^{d_3r}} \frac{1}{2} \|\FOmeg (\xb_t - \Ub\wb_t)\|^2_2$,} we find the partial derivatives of $\overline{\cL}_t$ with respect to $\Ub$:
\newthree{
\begin{align} \label{eq:grad:U}
\begin{split}
    \frac{\partial \overline{\cL}_t}{\partial \Ub} &= -\FOmeg'\FOmeg(\xb_t - \Ub\wb_t(\Ub))\wb_t(\Ub)'\\ &:= -\FOmeg'\FOmeg\rb_t\wb_t(\Ub)'.
\end{split}
\end{align}
}

\noindent \newthree{\newfour{See \cref{appendix:sec:proof_derivative} for the derivation of the gradient.} When computing the gradient, $\wb_t(\Ub)$ is solved for as detailed in the next subsection. }
{\color{black}
\begin{remark}
Note that the partial derivative \eqref{eq:grad:U} derived for the nested problem \eqref{eq:arb_missing:btd_stochastic_objective2} should not be confused with the case where $\overline{\cL}_t(\Ub)= \frac{1}{2}\|\FOmeg (\xb_t - \Ub \hat{\bmw}_t)\|^2_2$ for some fixed $\hat{\bmw}_t$. In that case, the gradient with respect to $\Ub$ is also $\frac{\partial \overline{\cL}_t}{\partial \Ub} =-\FOmeg'\FOmeg(\xb_t - \Ub\hat{\bmw}_t)\hat{\bmw}_t'$. However, 
the difference in our case is that, under the first assumption of \cref{prop:well_defined_tensor_grassmannian}, the weights $\wb_t(\Ub)= (\Ub' \FOmeg' \FOmeg \Ub)^{-1}\Ub' \FOmeg'$ are a function of  $\Ub$; See \cref{appendix:sec:proof_derivative} for further details.
\end{remark}
}


Using the work in \cite{edelman1998geometry}, the gradient \new{on the product of Grassmannians} in Fourier space is given by

\vspace{-3mm}
\begin{align}
    \nabla \overline{\cL}_t &= \mathcal{P_D}\left((\I - \Ub\,\Ub')\frac{\partial \overline{\cL}_t}{\partial \Ub}\right),
\end{align}
\vspace{-3mm}


\noindent where $\mathcal{P_D}(\cdot)$ sets the non-block-diagonal entries of the gradient to zero. The gradient of the objective on the product of Grassmannians then has the form \newthree{(using $\wb$ instead of $\wb_t(\Ub)$ for ease of notation and indexing the $d_3$ blocks of $\wb$)}

\vspace{-3mm}
\begin{equation}
    \nabla \overline{\cL}_t = \begin{bmatrix} -\new{\gammab_{1}} \wb_{1}' & & 0 \\ & \ddots & \\ 0 & &  -\new{\gammab_{d_3}} \wb_{d_3}'\end{bmatrix} \in \mathbb{C}^{d_1d_3 \times d_3r},
\end{equation}
\noindent where
\begin{align}
    \nabla \overline{\cL}_{t,k} &= -\new{\gammab_k} \wb_{k}' \in \mathbb{C}^{d_1 \times r} \label{eq:Ugrad:Lk}\\
    \new{\gammab_k} &= \left (\I - \Ub_k\Ub_k' \right)\rb_{{\Omega_t},k} \label{eq:Ugrad:rhob} \\
    \rb_{\Omega_t} &= \FOmeg'\FOmeg\rb_t = \texttt{unfold}(\texttt{fft}(\Delta_{\Omega_t}(\slice{R}_t),[],3))\label{eq:Ugrad:rbOmega}.
\end{align}
\vspace{-3mm}

\noindent Here, $\slice{R}_t = \new{\Delta_{\Omega_t}(\slice{X}_t)} - \slice{P}_t$, $\slice{P}_t = \tensor{U} \tP \slice{W}_t(\tU)$, $\Delta_{\Omega_t}(\cdot)$ imputes zeros on the unobserved tensor entries, and $\fft(\cdot, [], 3)$ takes the Fourier transform along the third-mode tubes. 


A gradient step along \new{each geodesic in the product manifold} with tangent vector $-\nabla \overline{\cL}_{t,k}$ is given by Equation (2.65) in \cite{edelman1998geometry} and is a function of the singular values and vectors of $\nabla \overline{\cL}_{t,k}$ \cite{grouse}. Each $\nabla \overline{\cL}_{t,k}$ has the rank-one SVD:

\vspace{-2mm}
\begin{align}
\label{eq:gradient_svds}
   \nabla \overline{\cL}_{t,k} &=  
    \begin{cases}
        \new{\bmu_k \sigma_k \bmv_k'}, & k = 1,\hdots, \lceil \frac{d_3 + 1}{2} \rceil \\
        \texttt{conj}(\nabla \overline{\cL}_{t,{(d_3 - k + 2)}}), & k = \lceil \frac{d_3 + 1}{2} \rceil + 1, \hdots, d_3
    \end{cases}\\
    &\new{\bmu_k = \frac{-\new{\gammab_k}}{\|\new{\gammab_k}\|}, \quad \bmv_k' = \frac{\wb_k'}{\|\wb_k\|} , \quad \sigma_k := \|\new{\gammab_k}\|\|\wb_k\|}. \nonumber
\end{align}


From \cite{edelman1998geometry}, a rank-one step of length $\eta > 0$ in the direction $-\nabla \overline{\cL}_{t,k}$ is given by
\begin{align} \label{eq:Ubar_rank_one_geodesic}
    &\Ub_{t+1,k} =
    \\ &\Ub_{t,k} + \left(\sin(\sigma_k \eta_k)\frac{\new{\gammab_k}}{\|\new{\gammab_k}\|} + (\cos(\sigma_k \eta_k) - 1)\frac{\pb_k}{\|\pb_k\|} \right) \frac{\wb'_k}{\|\wb_k\|}, \nonumber
\end{align}
where $\pb_k = \Ub_k \wb_k$ is the $k^{th}$ frontal face of $\overline{\tensor{P}}_t = \slice{P}_t \times_3 \FF_{d_3}$ (equivalently, in block diagonal matrix form, the $k^{th}$ block element of  \newthree{$\overline{\PP}_t = \Ub_t \Wb_t(\Ub_t)$, where $\Wb_t(\Ub_t)$} is the block-diagonal matrix formed from the $\wb_k$). Using conjugate symmetry of the Fourier transform, $\Ub_k = \texttt{conj}(\Ub_{(d_3 - k + 2)}), \quad k = \lceil \frac{d_3 + 1}{2} \rceil + 1, \hdots, d_3$.

Following from the result in \cite{grouse_greedy_step}, we use a greedy step size $\eta_k = \arctan(\|\new{\gammab_k}\|/\|\wb_k\|)$ on each Grassmannian. This choice of step size adaptively changes based on the FSM fit to the data, growing proportionally based on the angle between the projection and its residual. Using principles of conjugate symmetry of the FFT, we can save time by only computing the matrix-vector multiplications on half of the frontal slices in the Fourier domain and using the complex conjugate to find the others.

\subsection{Computing the weights \texorpdfstring{$\protect \slice{W}_t(\tU)$}{Wu}}
\label{subsec:w_update}

\newthree{For the gradient computations in $\Ub$, we first require computing}
\begin{align}
    \newthree{\wb_t(\Ub)} = \argmin_{\wb_t \in \mathbb{C}^{d_3r}} \frac{1}{2} \|\FOmeg (\xb_t - \Ub\wb_t)\|^2_2.
    \label{eq:w_objective}
\end{align}

If we were to solve for the optimal $\wb_t(\Ub)$ in closed form, this would require forming and inverting the $d_3r \times d_3r$ matrix $\Ub' \FOmeg' \FOmeg \Ub$, which can be very large. Instead, the block-wise separable structure of this quadratic problem suggests we use conjugate gradient descent (CGD) to estimate $\wb_t(\Ub)$ for a fixed $\Ub$. The structure permits fast, efficient computations by matrix-vector products $\Ub' \FOmeg' \FOmeg \Ub \vv$ for some vector $\vv \in \F^{d_3r}$. Within the t-SVD, this involves FFTs, separable matrix-vector products in each slice (from the block diagonal structure of $\Ub$), and zero-padding.

We observe faster overall convergence of our algorithm when solving the problem in $\wb_t(\Ub)$ at each time step with higher accuracy. CGD is guaranteed to converge in as many iterations as the dimension of the optimized vector \cite{shewchuk1994introduction}, but since $\wb_t(\Ub)$ is $d_3r$-dimensional, the number of maximum iterations could be rather large. As the number of missing entries increases, the matrix $\FOmeg\Ub$ in the least squares problem of Eq. (\ref{eq:w_objective}) becomes more poorly conditioned, and since the convergence rate of CGD is dependent on the condition number of this matrix, denoted $\kappa(\FOmeg\Ub)$, the algorithm will require more iterations to solve the problem to within some $\epsilon > 0$ accuracy, slowing the run-time of our algorithm. However, as noted above, it is impractical to form and store the large matrix, much less compute its SVD to find $\kappa$. We prove a practical upper bound on the number of CGD iterations as a function of the sampling rate and show CGD converges in far fewer iterations than the maximum for most subsampling rates. The proof, along with accompanying lemmas, is left to \cref{proofs:cgd}. Empirical studies of our algorithm show the number of maximum CGD iterations is tightly bounded by our \cref{thm:sample_bound_coherence} for most subsampling rates.

\new{
For the following theorem, we will require a notion of tensor coherence, given in \cite{Zhang_Aeron}:
\begin{definition}[Tensor coherence]\label{defn:cohe} 

Let $\tensor{U} \in \mathbb{R}^{d_1 \times r \times d_3}$ be an orthonormal tensor whose $r$ lateral slices span the free submodule of $\mathbb{M}^{d_1}_{d_3}$.  Then, the $\mu$-coherence of $\tU$ is given by
    \begin{align}
        \mu(\tU) := \max_{i=1,\hdots,d_1} \|\tU ^T \tP \mathring{\bme}_i\|_2^2,
    \end{align}
    where $\mathring{\bme}_i \in \R^{d_1 \times 1 \times d_3}$ is the column basis with $\mathring{\bme}_{i11} = 1$ and the rest of the entries are zero. Note that $\frac{r}{d_1 d_3} \leq \mu(\tU) \leq 1$.
\end{definition}

\new{It is standard practice in matrix and tensor completion literature to make some assumption that the coherence is not too large to guarantee recovery (see Candes \& Recht \cite{candes_recht} for matrix completion and Zhang \& Aeron \cite{Zhang_Aeron} who consider tensor completion under the t-product.) We will impose a coherence upper bound assumption on all of the iterates of $\tU_t$ as well as a sampling condition for the number of entries per slice that must be observed.}
\begin{theorem}
    Let $\PP_{\Omega_t}$ sample $|\Omega_t|$ rows from $(\FF_{d_3}' \otimes \I_{d_1})\Ub$ uniformly at random such that $|\Omega_t|/\log(|\Omega_t|) > C^2 \mu_0 d_3r$, where $C$ is a universal constant and $\mu_0 > 1/d_3$ is small. Assume coherence of the $\tU_t$ iterates remains bounded as $\mu(\tU_t) \leq \frac{\mu_0 r}{d_1}$. Then with probability at least $1 - \delta$, where $\delta \in [0,1]$, the maximum number of conjugate gradient descent iterations, $J$, required to solve \cref{eq:w_objective} to within $\epsilon$-precision for $\epsilon > 0$ is upper bounded as:
       \begin{align}
           &J \leq \frac{1}{2} \sqrt{\frac{1 + \delta^{-1}\tau}{1 - \delta^{-1}\tau}} \log(2/\epsilon),\\
           &\text{ where } \tau = C\sqrt{\mu_0 r d_3 \frac{\log(|\Omega_t|)}{|\Omega_t|}}. \nonumber
       \end{align}
      \label{thm:sample_bound_coherence}
\end{theorem}
The proof is found in \cref{proofs:cgd}.}

\begin{table*}[!t]
\centering
\setlength{\tabcolsep}{7pt}
\begin{tabular}{|c|c|c|c|}
\hline
{Algorithm}  & {Memory per iteration}  & {Total batch computational complexity}  & {Computation per iteration}\\
\thickhline

TOUCAN & $O(d_1rd_3 + rd_3)$ & $O(B(J(d_1Td_3\log(d_3) +  d_1rTN)))$ & $O(J(d_1d_3\log(d_3) +  d_1rN))$\\
TCTF & $O(d_1rd_3 + rTd_3)$ & $O(A(d_1Td_3\log(d_3) + d_1rTN))$ & $O(d_1Td_3\log(d_3) + d_1rTN)$\\
TNN-ADMM & $O(d_1Td_3)$ & $O(A(d_1Td_3\log(d_3) + d_1TN\min(d_1,T)))$ & $O(d_1Td_3\log(d_3) + d_1TN\min(d_1,T))$ \\
 
\thickhline
\end{tabular}
\caption{\new{Algorithm memory and computational complexities. $A$ here denotes the number of algorithm iterations for batch methods, and $B$ denotes the number of batch passes for TOUCAN. Here usually $B\ll A$.}}
\label{tab:complexity}
\end{table*}

\subsection{Algorithm}
\label{subsec:algorithm}
The preceding updates give an efficient algorithm we call TOUCAN (Tensor rank-One Update on the Complex grassmanniAN) for computing each variable in the Fourier domain with simple, efficient linear algebra operations and fast Fourier transforms. TOUCAN is numerically stable by maintaining orthonormality on the product of Grassmannians and is constant in memory use, scaling linearly with the number of observed data samples instead of in polynomial-time like batch t-SVD methods. \new{In addition, like other t-SVD algorithms, independent computations in each slice (equivalently the blocks of the block-matrix terms) in the Fourier domain can be carried out in parallel.} TOUCAN is summarized in Algorithm \ref{alg:toucan}.

\new{ TOUCAN can handle two cases of online and streaming data. The first is incremental batch completion where the batch tensor is too large to read into local memory, but can be stored elsewhere. Our algorithm reads only slice $\slice{X}_t$ into local memory, updates its estimate of $\tU$ and weights $\slice{W}_t(\tU)$, discards this local copy of $\slice{X}_t$, and passes over each data slice like this in sequence. In this setting, it is possible to make multiple passes over the full batch while only reading parts into memory. This is a sensible approach when the underlying low-rank model is believed to be static or stationary throughout the batch. The second use case of TOUCAN is for purely streaming data where we seek to learn $\tU$ from each new observation and discard each observation completely after processing. The algorithm then tracks any changes in $\tU$ only from new observations and is able to track a time-varying low-rank model. }
    

\begin{algorithm}
\begin{algorithmic}[1]

\REQUIRE{\textbf{Data:} $\slice{X}_t \in \mathbb{R}^{d_1 \times 1 \times d_3} \quad \forall i=1,\hdots,T$ observed on $\Omega_t$; tubal-rank $r$, tolerance $\epsilon >0$.}
\STATE Initalize Fourier transformed orthonormal tensor $\tUb_0 \in \mathbb{C}^{d_1 \times r \times d_3}$.
\FOR{$t=1$ to $T$}
\STATE  Compute $\tXb_{\Omega_t} = \new{\texttt{fft}(\Delta_{\Omega_t}(\tensor{X}_{\Omega_t})},[],3)$.

\STATE Estimate \newthree{optimal weights $\wb_t(\Ub)$} by solving Eq. (\ref{eq:w_objective}) with CGD to within tolerance $\epsilon >0$.

\STATE {Predict full slice in the Fourier domain:  ${\overline{\PP}_t = \Ub_t {\color{black}\Wb_t( \Ub_{t})}}$.}

\STATE {Shape into tensor and transform:  ${\slice{P}_t = \texttt{ifft}(\overline{\tensor{P}}_t,[],3)}$.}

\STATE {Compute residual: } $\slice{R}_t = \new{\Delta_{\Omega_t}(\slice{X}_t)} - \slice{P}_t$.

\STATE {Compute gradient terms} from Eqs. (\ref{eq:Ugrad:rhob}) and (\ref{eq:Ugrad:rbOmega}).


\STATE {Update subspace: } $\tUb_{t+1}$ from \cref{eq:Ubar_rank_one_geodesic}.

\STATE Transform: $\tensor{U}_{t+1} = \texttt{ifft}(\tUb_{t+1},[],3)$.
\STATE Transform: ${\color{black}\slice{W}_t (\tensor{U}_{t})} = \texttt{ifft}({\color{black}\tWb_t (\tUb_{t})}  ,[],3)$.

\ENDFOR

\RETURN{} $\tensor{U}, \slice{W}_t(\tU_t), \quad \forall t=1,\hdots,T$

\end{algorithmic}
\caption{Tensor rank-One Update on the Complex grassmanniAN (TOUCAN): Arbitrary Missing Tensor Entries}
\label{alg:toucan}
\end{algorithm}

  
   


\subsection{Memory and computational analysis} TOUCAN processes a tensor incrementally and thus only needs to store an orthonormal tensor $\tU \in \R^{d_1 \times r \times d_3}$, the weights $\slice{W}_t(\tU) \in \R^{r \times 1 \times d_3}$ per $t=1,\hdots,T$, requiring $d_1d_3r + d_3r$ memory elements per \new{iteration at time $t$}. Upon updating $\tU$, the new $\slice{W}_{t+1}(\tU)$ is computed at the next iteration using the same memory. At each time instance, this is far less than storing the entire tensor in memory which would require $d_1Td_3$ memory elements, especially when any of the dimensions is very large. 

Implemented efficiently, the main loop of our algorithm requires 4 fast inverse Fourier transforms and one fast Fourier Transform. \sloppy The CGD update takes $O(J(Nd_1r + d_1d_3\log(d_3))$ flops where $N = \lceil (d_3 + 1)/2 \rceil$ and $J$ is the number of CGD iterations. Computing $\slice{R}$ takes $O(Nd_1r + d_1d_3\log(d_3) + d_1d_3)$ flops. The update in \cref{eq:Ugrad:rbOmega} takes $O(d_1d_3\log (d_3))$ flops, and \cref{eq:Ugrad:rhob} takes $O(Nd_1r)$ flops. Computing the subspace update requires $O(Nd_1r)$ flops. \new{\cref{tab:complexity} summarizes the memory and computational requirements of our algorithm compared to other t-SVD algorithms.}

\subsection{Convergence}
\label{subsec:convergence}
\new{

\newnew{
 Here, we prove expected linear local convergence of our algorithm. Our analysis follows naturally from the work in \cite{balzano_wright2015}, which proved a similar result for the GROUSE algorithm, for which TOUCAN is a related extension to t-product tensors. The problem of proving convergence for this class of algorithms is complicated by the setting of streaming data, missing entries, and the optimization problem being constrained to a nonlinear manifold with rank-one updates. Other analyses for related problems may exist like in \cite{huang2021robust}, but often these are limited to the cases of batch data, fully observed entries, and a specific type of retraction operator on the manifold. However, little work in the literature exists for finite-sample analysis with missing data, and the only results for streaming subspace estimation with missing data in the matrix case are local convergence results for the GROUSE algorithm in \cite{balzano_wright2015} and the work in \cite{de2015global}; see \cite{balzano2018streaming} for a recent survey of the area.
 
 Our problem and analysis are tailored to these specific settings, and we apply similar assumptions as made in \cite{balzano_wright2015} and other standard assumptions made in the literature. For simplicity of analysis, we focus on the case of tubes sampled uniformly at random since it allows us to extend the results provided in \cite{balzano_wright2015} to the tensor case.

Our theory provides expected linear local convergence under (i)
the randomness of the observed tensor and (ii) the randomness of the subset of elements observed at each iteration. More specifically, we have the following assumptions:


\begin{assumption}{1}{}\label{A1_article}
Each $\slice{X}_t = \tU^* \tP \slice{S}_t$ for planted model $[\tU^*] \in \mathcal{G}(r,d_1,d_3)$ and $(\slice{S}_t)_{ijk} \stackrel{i.i.d.}{\sim} \mathcal{N}(0,1)$. 
\end{assumption}


\begin{assumption}{2}{}\label{A2_article}
Let $\Omega_t \subset \{1,\hdots,d_1\}$ denote the tube indices, and assume the tube indices are chosen uniformly at random. In other words, the data $\slice{X}_t $ are sampled tubal-wise, where a tube consists of $d_3$ entries along the third mode dimension.
\end{assumption}

It is worth mentioning that \ref{A1_article} is a generalization of the assumption made in \cite{balzano_wright2015} to the tensor problem. We also note that conditions analogous to \ref{A2_article} have been used in \cite{liu2015adaptive} for t-product based tensor completion tasks. 

Before providing the main result, we give some additional notations and definitions.  Let $\mu(\Ub)$  denote the matrix coherence of complex orthonormal matrix $\Ub \in \mathbb{C}^{d \times r}$, i.e. 
$\mu(\Ub) := \frac{d}{r} \max_{i=1,\hdots,d} \|\Ub'\bme_i\|_2^2$, where $\bme_i$ is the $i^{th}$ standard basis unit vector in $\mathbb{C}^{d}$. We note this is consistent with Definition \ref{defn:cohe}. For a vector argument $\xb \in \C^d$, this further specializes to $\mu(\xb) = d \|\xb\|_\infty^2 / \|\xb\|_2^2$. We use $\mu_{\max}(\tU^*) := \max_{k=1,\hdots,d_3} \mu(\Ub^*_k)$ to denote the maximum coherence of the frontal slices of $\tUb$.

Denote
\begin{equation}\label{eqn:epsi:block_article}
 \epsilon_{t,k} := r - \|{\Ub_k^*}'\Ub_{t,k}\|_F^2, \quad \forall k \in [d_3].
\end{equation}

We will analyze the sequence $\epsilon_{t}$ measuring the error between the planted model $\tU^*$ and the algorithm's estimate $\tU_t$: 
\begin{align}\label{eq:epsilon_def_article}
\nonumber
    \epsilon_t &:= d_3r - \|{\tU^*}' \tP \tU_t\|_F^2 \\
     &= d_3r - \|{\Ub^*}'\Ub_t\|_F^2  = \sum_{k=1}^{d_3} \epsilon_{t,k},
\end{align}
 where $\Ub^* (\text{resp. } \Ub_t)$ denotes the current estimate of the block-diagonal Fourier representation of $\tU^* (\text{resp. } \tU_t)$. Again, here we use the normalized DFT matrix when taking the Fourier transform. The second equality follows from the definition of the Frobenius norm under the t-product. If $\tU_t$ perfectly estimates the free submodule spanned by $\tU^*$, it's easy to see from \cref{eq:epsilon_def_article} that $\epsilon_t = 0$; on the other hand, if $\tU_t$ is orthogonal to $\tU^*$ in t-product, $\epsilon_t = d_3r$. 
 
\begin{theorem}\label{thm:local_article}
Let $\{({\color{black}\slice{W}_{t-1} (\tU_{t-1})},\tU_t )\}_{t\geq 1}$ denote the sequence generated by Algorithm~\ref{alg:toucan_missing_tubes}. Suppose \ref{A1_article} and \ref{A2_article} hold, and the number of sampled tubes $|\Omega_{t}| \geq q$ for all $t$ such that 
  \begin{equation}\label{eqn:samb_article}
 q \geq C_1 \log(d_1)^2 r \mu_{\max}(\tU^*) \log(20rd_3)     
  \end{equation}
for some $C_1\geq 64/3$. Suppose there exists $\bar{\delta} \in (0, 0.6/d_3)$ such that the residual vector $\vb_{t,k} :=\xb_{t,k} - \Ub_{t,k} \Ub_{t,k}' \xb_{t,k}$ satisfies
\begin{subequations}\label{equ:mu:coh_article}
  \begin{align}
  \mu(\vb_{t,k}) &\leq \log d_1  \left(\frac{0.045}{\log(10d_3)} C_1 r \mu(\Ub_{t,k}) \log(20rd_3) \right)^{\frac{1}{2}},  \label{eq:resid1_article}\\
  \mu(\vb_{t,k}) &\leq (\log d_1)^2  \frac{0.05}{8\log (10d_3)} C_1  \log(20rd_3), \label{eq:resid2_article}
  \end{align}
\end{subequations}
for all $k=1,\hdots,d_3$ each with probability at least $1-\bar{\delta}$. Assume further that 
\begin{subequations}
\begin{align}
  \epsilon_{t,k} &\leq \min \left\{\frac{q^2}{128 d_1^2 r},\frac{r}{16d_1}\mu(\Ub^*_k)\right\}, ~\forall k\in[d_3],
  \label{eqn:epsi_article} \\
    \epsilon_{t}  &\leq (8 \cdot 10^{-6})(0.6-d_3\bar{\delta})^2\frac{q^3}{d_1^3r^2}. \label{eqn:epsi:total_article}
\end{align}
\end{subequations}
Then, 
  \begin{align}\label{eqn:fin:rate_article}
\mathbb{E}[\epsilon_{t+1}|\epsilon_t] \leq \left(1 - 0.16(0.6  - d_3\bar{\delta})\frac{q}{d_1r}\right)\epsilon_t.
  \end{align}
\end{theorem}

\begin{remark}
We note that in the matrix case, i.e., $d_3=1$, our result recovers \cite[Corollary~2.15]{balzano_wright2015}. We also note that the failure probability increases as $d_3$ grows which is similar to the results provided in the t-SVD literature under the tubal sampling assumption; see, e.g., \cite[Theorem 3.2]{Zhang_Aeron}. 
\end{remark}

\begin{remark}
The analogous supposition that Equations \eqref{eq:resid1_article} and \eqref{eq:resid2_article} hold was also made in \cite{balzano_wright2015} for the matrix case, where they pointed out that empirical evidence supports this assumption. This is essentially assuming that the residual vectors are roughly as incoherent as the subspaces themselves. 
The assumptions in \eqref{eqn:epsi_article} and \eqref{eqn:epsi:total_article} define the local region in which the expected linear rate of convergence is achieved. As discussed in \cite{balzano_wright2015}, this local region is conservative according to empirical evidence, which is also supported by the experiments in our own work. 
\end{remark}

\begin{remark}
In \eqref{eqn:fin:rate_article}, the expectation is with respect to the randomness in the data drawn from $\tU^*$ with normally distributed coefficients, and the high probability result is with respect to the tubes observed at random. Supposing that we are within the radius of local convergence, the rate \eqref{eqn:fin:rate_article} suggests our algorithm converges faster the closer $\frac{q}{d_1r}$ is to 1, and the fastest when we observe fully sampled data. Conversely, with fewer tube observations $q$, the rate of convergence slows. 
\end{remark}

}
}


\section{Experimental Results}
\label{sec:experiments}

\subsection{Numerical experiments}
\label{subsec:synthetic}
\begin{figure*}[!t]
    \centering

    \subfloat[$r=3$, $d_1=100$, $d_2=1000$, $d_3=20$, $\sigma^2=0.$]{\includegraphics[width=0.23\textwidth,trim={0 0.5cm 0 0cm},clip]{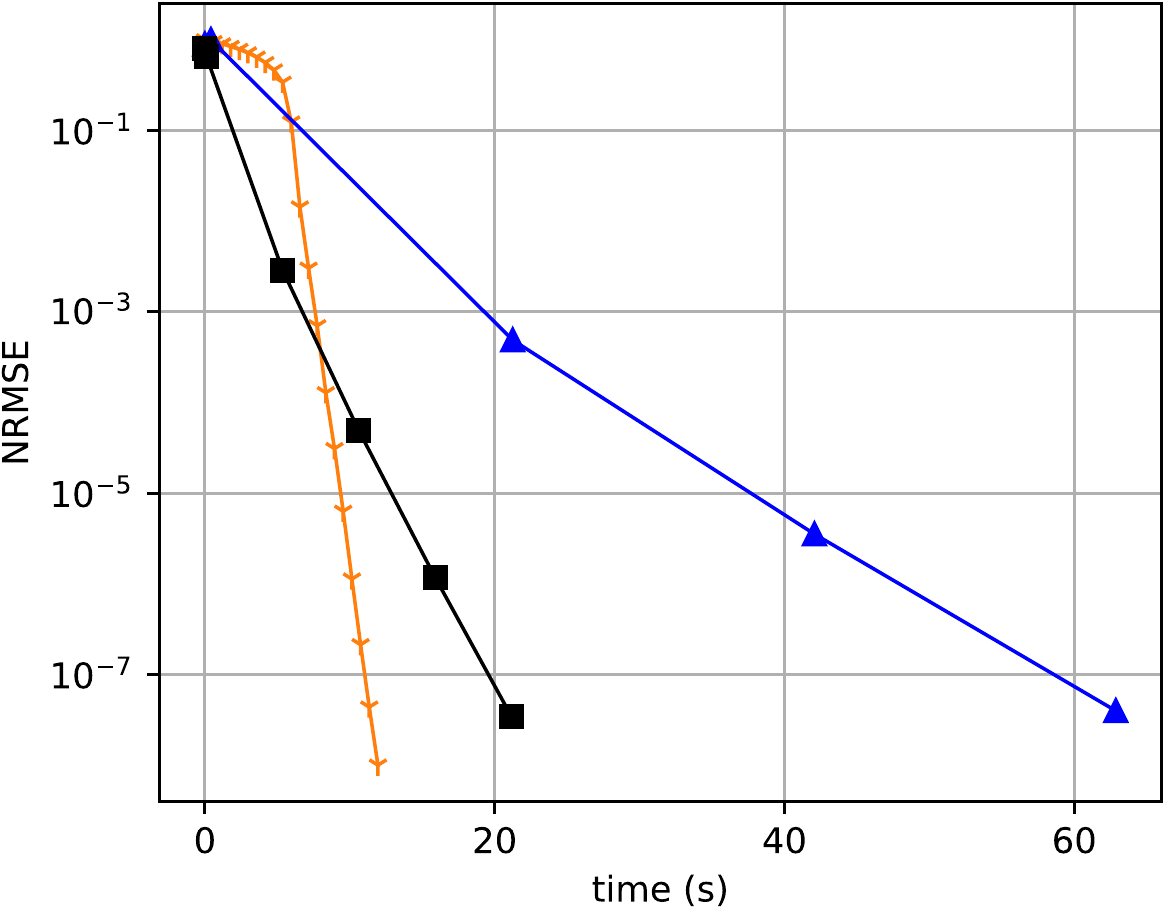}}
    \hfill
   \subfloat[ $r=3$, $d_1=100$, $d_2=1000$, $d_3=20$,  $\sigma^2=10^{-3}.$]{\includegraphics[width=0.23\textwidth,trim={0 0.5cm 0 0cm},clip]{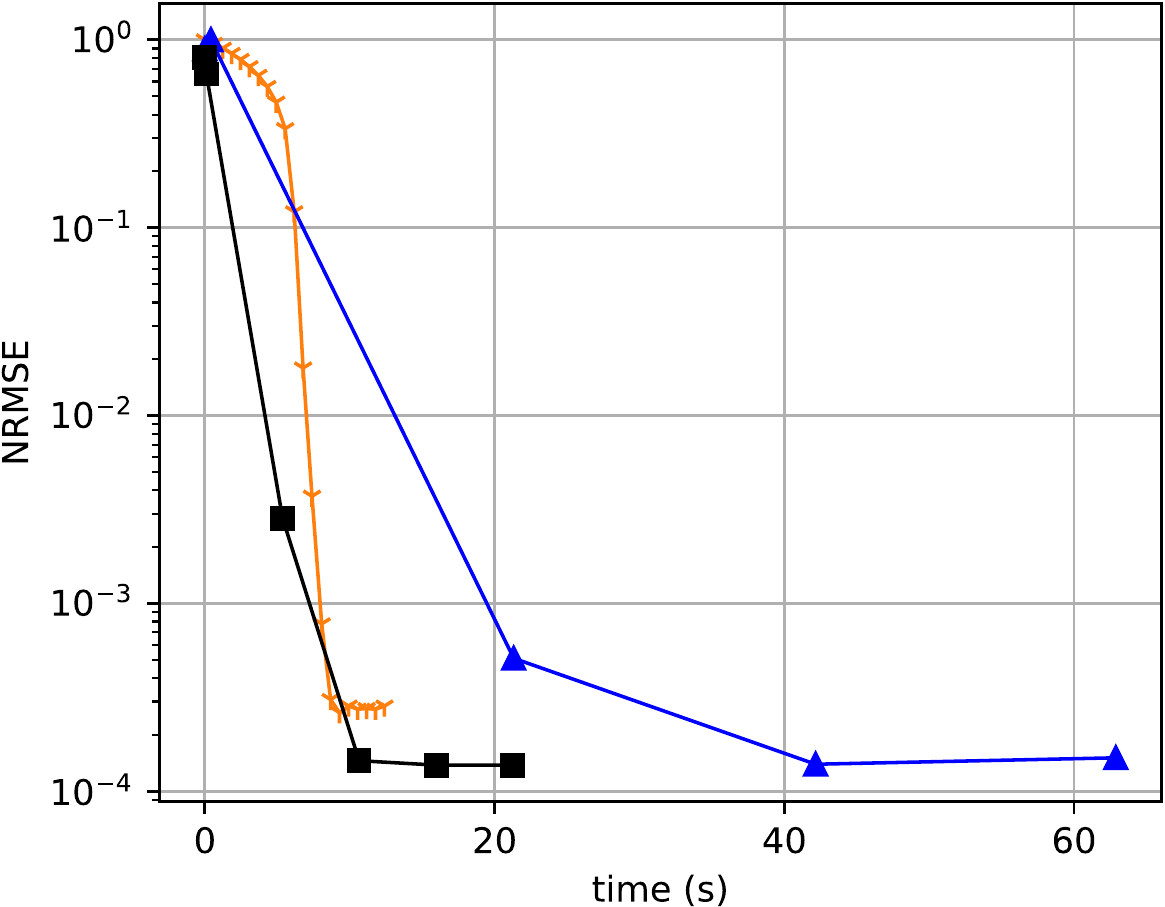}}
    \hfill
    \subfloat[ $r=3$, $d_1=200$, $d_2=1500$, $d_3=50$, $\sigma^2=0.$]{\includegraphics[width=0.23\textwidth,trim={0 0.5cm 0 0cm},clip]{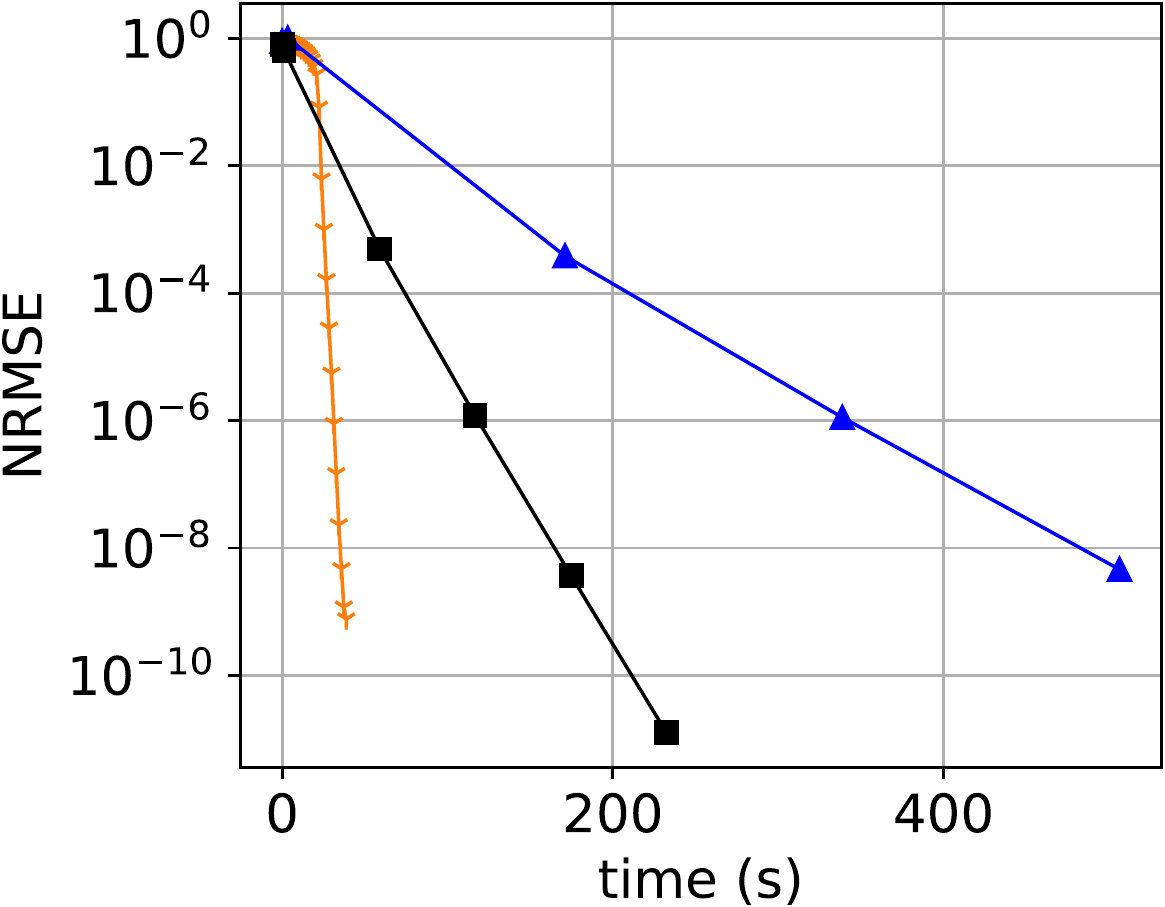}}
    \hfill
    \subfloat[ $r=3$, $d_1=200$, $d_2=1500$, $d_3=50$, $\sigma^2=10^{-3}.$]{\includegraphics[width=0.23\textwidth,trim={0 0.5cm 0 0cm},clip]{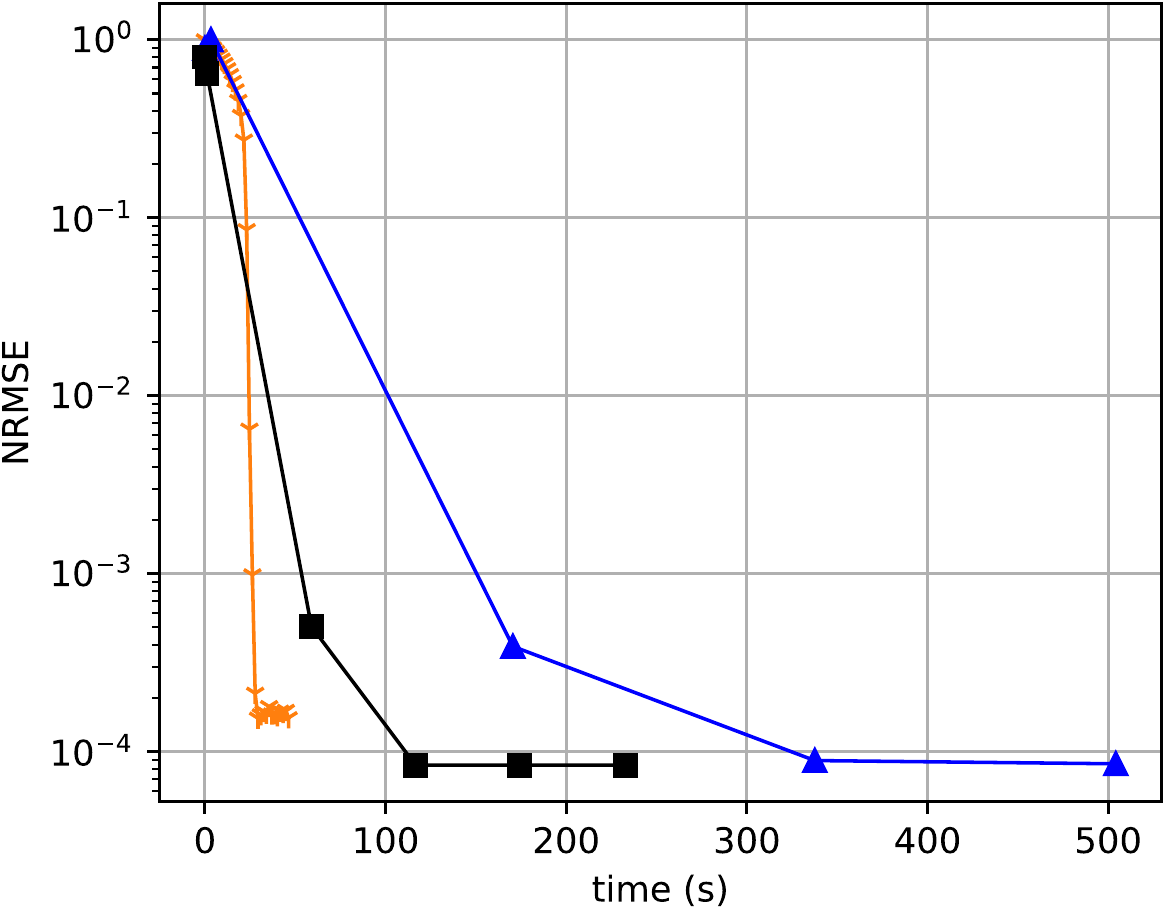}}
    \\
     \subfloat[$r = 3, d_1 = d_3 = 100$, $\sigma^2 = 0$.]{\includegraphics[scale=0.3,trim={0 0 6.6cm 0},clip]{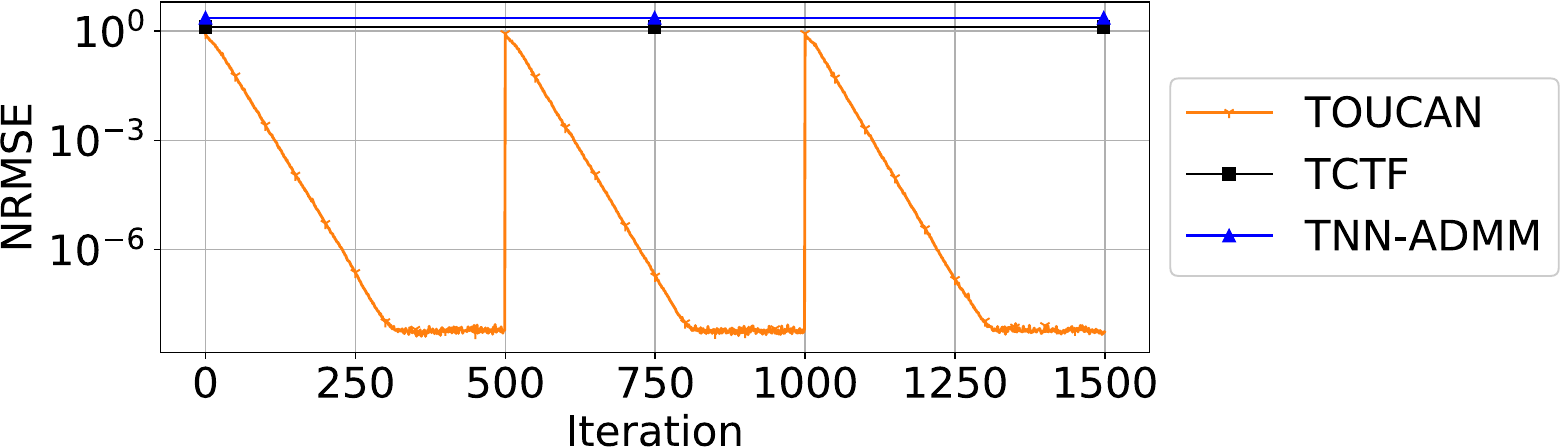}}
    %
    \subfloat[$r = 3, d_1 = d_3 = 100$, $\sigma^2 = 10^{-3}$.]{\includegraphics[scale=0.3]{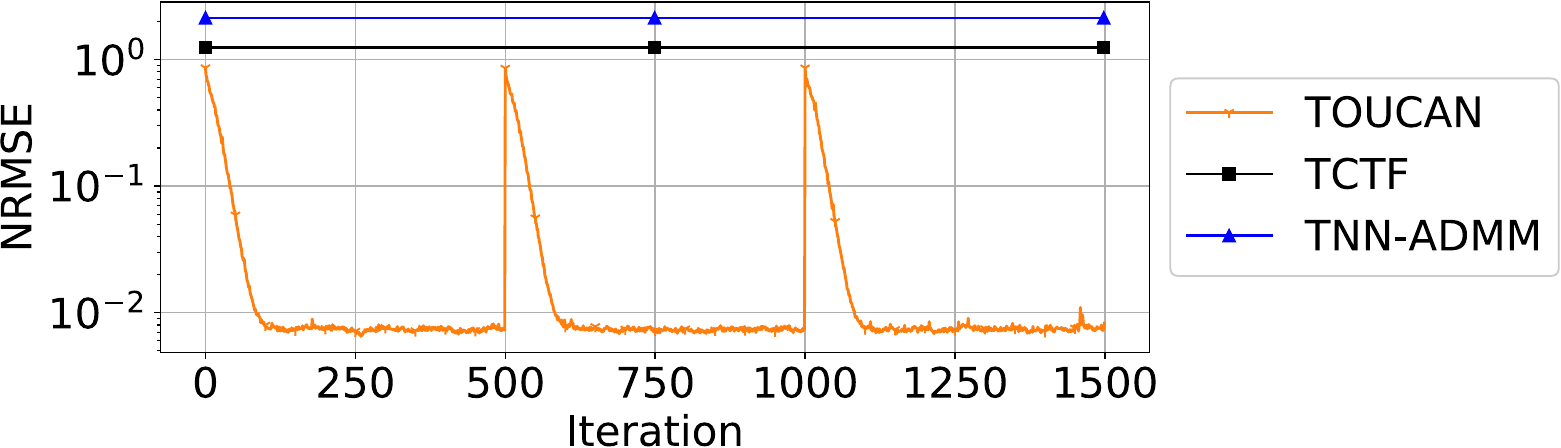}}
    
    \caption{(a)-(d): Batch completion of t-SVD synthetic tensors with 20\% entries observed and median wall-clock time over 10 trials on the $x$-axis. Markers are plotted every 100 TOUCAN iterations and 50 batch algorithm iterations. (e) \& (f): TOUCAN completing a tensor from a dynamically changing FSM over time compared to batch completion t-SVD methods with 50\% of the entries observed. Markers are plotted for every 50 TOUCAN iterations. \new{The second dimension of the tensor passed to the batch algorithms TCTF and TNN-ADMM is equal to the number of iterations.}}
    \label{fig:synth_batch}
\end{figure*}
\subsubsection{Incremental Tensor Completion}
\label{subsec:inc_tensor_comp}
We first verify the validity and efficiency of TOUCAN in recovering large-scale missing tensor data from synthetically generated isotropic Gaussian distributions with low-tubal-rank. We compute the t-product of two low-tubal-rank tensors $\tensor{U} \tP \tensor{W}$ to yield a third order tensor of tubal-rank $r=3$ and sample 20\% of tensor entries/tubes randomly according to a Bernoulli distribution. TOUCAN observes one lateral slice sequentially, solves the inner CGD step to within a set tolerance ($10^{-6}$), and is allowed to process over the entire batch twice. Our simulations compare against t-SVD batch tensor completion algorithms: 1) an algorithm that optimizes tensor nuclear norm via alternating direction method of multipliers (TNN-ADMM) \cite{Zhang_Aeron}, and 2) Tensor Completion by Tensor Factorization (TCTF) of \cite{tctf_algorithm}, which factorizes the tensor for the t-product of two low-tubal-rank tensors. \new{For TCTF, we omit the rank-reduction steps and set \newnew{the multi-ranks equal to the planted tubal-rank since our synthetic examples are generated in this manner}, and the steps only add computation; this also makes TCTF more comparable to TOUCAN since they both seek a similar nonconvex factorization under the t-product.} \cref{fig:synth_batch}(a)-(d) plot the normalized root-mean-squared error (NRMSE) $\|\tX_\text{est.} - \tX_\text{true}\|_F / \|\tX_\text{true}\|_F$ of the recovered tensor to the true tensor by median elapsed wall clock time in seconds over 10 trials. In addition, we also examine cases with additive white Gaussian noise. \newnew{All algorithms are coded in Python with our optimized implementations of TNN-ADMM, TCTF, and STC; for OLSTEC and TeCPSGD, we use the implementations from \cite{olstec} converted to Python.} Experiments were run on a Intel(R) Core(TM) i7-6850K CPU @ 3.60GHz. Our implementation can be found at \url{https://github.com/kgilman/TOUCAN}. 

For tensors with large $d_2$ dimension, TOUCAN can rapidly complete the data in substantially less time than either batch algorithm while using only 0.3\% of the memory per iteration compared to storing the entire tensor for our synthetic example. \cref{fig:synth_batch} shows the algorithm scales up well with the tensor dimensions, and can achieve batch completion for large-scale tensors in orders of magnitude less computation time. With additive Gaussian noise, our stochastic gradient algorithm achieves accuracy to within a noise floor proportional to the noise variance. For smaller size tensors, the batch algorithms succeed in less wall clock time, so TOUCAN only becomes advantageous when the dimensions of the tensor scale to be very large. With larger amounts of additive noise, the algorithm's advantage diminishes since the batch algorithms are able to more quickly average the noise out than our stochastic algorithm. \new{Since the first observations in the initial phase of the algorithm will be revisited in later passes over the batch, the NRMSE curve with respect to the entire tensor shows slower progress for TOUCAN in the first iterations, but this graph under-reports the accuracy of the estimate of $\tU$.}

\subsubsection{Dynamic FSM Tracking}
We demonstrate TOUCAN's ability to track a dynamically changing FSM from streaming multiway data with missing entries. We generate a random orthonormal basis $\tensor{U}$ for various tubal-ranks from an i.i.d. Gaussian distribution and draw 2-D lateral slices by t-product with i.i.d. Gaussian weights. 50\% of the tensor entries are sampled at random, and we record the NRMSE of the completed tensor slice $\|\slice{X}_{t,\text{est.}} - \slice{X}_{t,\text{true}}\|_F / \|\slice{X}_{t,\text{true}}\|_F$. The experiment simulates abrupt system dynamics by randomly reinitializing the underlying FSM every 500 slices. The results in Fig. \ref{fig:synth_batch}(e),(f) illustrate TOUCAN's ability to adaptively re-estimate each new FSM and capture system dynamics that the batch algorithms cannot, as they compute estimates based on the entire batch of data collected over time.

    


\subsection{Real data experiments}
\label{subsec:experiments_real_data}

\subsubsection{Application to Gas Measurements Tensor}
\label{subsec:experiments:gas}
We deploy our algorithm to track a dynamically changing free submodule from streaming 2-D lateral slice data with missing entries in chemo-sensing data collected by Vergara et al.~\cite{vergara}. The dataset consists of measurements as a gas is blown over an array of conductometric metal-oxide sensors in a wind tunnel \cite{kolda_hong}. The data is made up of six arrays each with 72 sensors, 260 seconds of data points collected at $\sim100$ Hz, and 300 experiments for each of 11 gases.  The sensor values vary in time as a gas permeates throughout a wind tunnel and then dissipates \cite{kolda_hong}.  We chose to fix the array and gas, using the fourth sensor array and Toulene gas for our experiments, downsample to 10 Hz, and remove sensor 33 (out of 72) and time samples 1103 and 2012, which seemed to have erratic measurements, resulting in a tensor of size $300\times2600\times 71$. We subtract the sample mean from the columns of each time slice -- a column referring to 300 experiment samples per sensor -- normalize each time slice by its Frobenius norm, and subsample only 25\% of the data to simulate missing entries.

TOUCAN is compared to the batch t-SVD algorithms, the two online CP algorithms TeCPSGD and OLSTEC (we use the source code from \cite{olstec} for our implementations of OLSTEC and TeCPSGD), and the online tensor completion algorithm Sequential Tensor Completion \cite{stc}, which estimates an orthonormal rank-$r$ unfolding for each mode, to recover undersampled chemo-sensing data. The online algorithms process each time slice sequentially, observing a $300 \times 71$ matrix of experiments versus sensor channels, and pass over the entire data once. We empirically found tracking a $1$-dimensional FSM with TOUCAN to have the best performance. The algorithm updates its estimate of $\tensor{U}_{t+1}$, and weights {\color{black}$\slice{W}_{t} (\tensor{U}_{t})$}.

For competing algorithms, we tuned parameters by grid search (see \cref{sec:hyperparams}) to find the best performance in NRMSE on the first 300 time samples. The online CP algorithms learn a rank-50 decomposition, \new{with their factors initialized with the left-singular subspaces of the mode unfoldings of the first 300 samples.} We set $\lambda = 10^{-5}$ and the initial step size to be $10^3$ for TeCPSGD, and $\lambda = 0.9$ and $\mu = 10^{-8}$ for OLSTEC. STC learns a multirank (15, 15, 1) model. TCTF learns a tubal-rank 10 factorization, and the ADMM algorithm penalty is set to be $\rho = 1.5$ for TNN-ADMM. The batch algorithms iterate until the difference NRMSE between iterates is less than $10^{-4}$ or a maximum number of 75 iterations is reached. 

\begin{figure}
    \centering
    \includegraphics[width=0.48\textwidth]{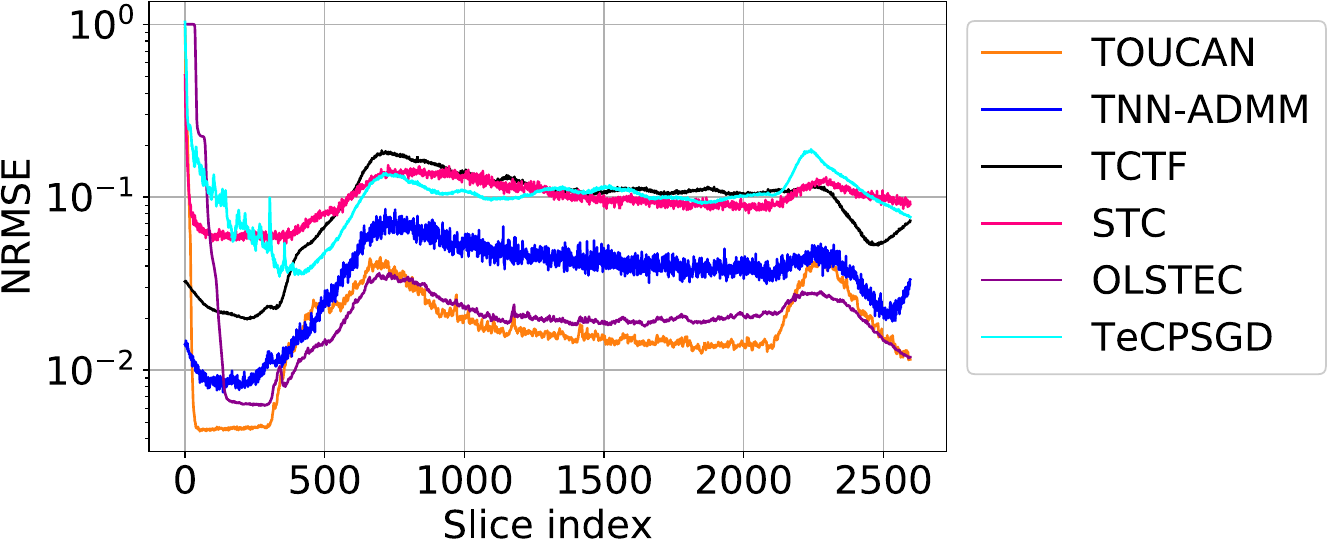}
    \captionof{figure}{NRMSE of each recovered time slice for Toluene gas dataset from 25\% samples.}
    \label{fig:gas_nrmse}
\end{figure}

\begin{table}
\caption{Total wall clock times in seconds for Toluene gas dataset}
    \centering
    \begin{tabular}{c c}
    \hline
    \textbf{Algorithm} & \textbf{Time (s)} \\
    \thickhline
    TOUCAN & 30.81 \\
    TeCPSGD & 137.31\\
    STC & 363.54 \\
    OLSTEC & 552.53\\
    TCTF & 725.55\\
    TNN-ADMM & 879.78 \\
    \thickhline
    \end{tabular}
    \label{tab:gas_times}
\end{table}

Fig. \ref{fig:gas_nrmse} compares the NRMSE of each recovered 2D slice to the true data at each time instance for the algorithms, which shows TOUCAN tracking the sensor readings with comparable error to OLSTEC. \new{Due to the non-stationary behavior of the data, the tracking errors fluctuate as the data changes in time. While the batch methods achieve the best overall NRMSE error computed for the entire tensor, the online methods show the best reconstruction error on each sample after the initial start-up iterations.} We also give the total computation time for each algorithm in \cref{tab:gas_times}, emphasizing the significant speedup TOUCAN attains over the baseline algorithms, particularly the batch algorithms that are computationally prohibitive with large tensor data.

\subsubsection{Streaming dynamic MRI reconstruction}
\label{subsec:experiments:mri}
Magnetic resonance imaging (MRI) collects a high-dimensional tensor that is often undersampled due to computational limitations exacerbated by large volumetric and dynamic acquisitions. One successful solution to image reconstruction from limited sampling is low-rank tensor completion \cite{banco_aeron,mardani}. A t-SVD factorization of the spatial frequency-by-time (or $k$-$t$ space) tensor reveals low-tubal-rank structure in the real and complex components \cite{banco_aeron}, and t-SVD algorithms have been shown to be proficient at completing the $k$-$t$ space tensor for image reconstruction. MRI data can also contain significant motion content and time-varying dynamics such as breathing motion. We employ TOUCAN's ability to track streaming time-dynamic multiway day to recover the $k$-$t$ space tensor.

We test the completion abilities of each algorithm on the invivo myocardial perfusion dataset data from \cite{lingala_et_al} with both varying levels of uniformly random entry sampling and tube sampling along the $k_y$ direction. 
The dimensions of the data are $k_x = 190, k_y = 90$ and $k_t = 70$, and the data contains many dynamic motions such as heartbeats, breathing motion, and image intensity changes. 

The streaming algorithms pass over the data once with the $k$-space rows oriented along the third tensor mode ($k_y = d_3$). TOUCAN learns a free submodule of tubal-rank 5, and two streaming CP algorithms learn a rank-50 CP decomposition.
\new{After exhaustive search for hyperparameters,} we set $\lambda = 0.5$ and $\mu=10^{-4}$ for OLSTEC, and $\lambda=10^{-4}$ and the initial step size to be $10^5$ for TeCPSGD. We set the ranks to be $r_1 = r_2 = 25, r_3 = 5$ for STC. STC cannot handle tube-sampled data since an entire column of one of the tensor unfoldings will be missing, so we only test it in the case where arbitrarily random entries are missing. The batch t-SVD algorithms are allowed to compute over the data until the difference in NRMSE between iterates is less than $10^{-4}$ or the algorithm exceeds a specified maximum number of iterations. 

We record the NRMSE, mean structural similarity index measures (SSIM) \cite{ssim} of the reconstructed images, and total algorithm wall-clock times in Table \ref{tab:mri_stats_cardiac}. Fig. \ref{fig:cardiac_reconstruction_40p} displays a sample of the reconstruction results, along with plots of the NRMSE of each frame's recovered real $k$-$t$ space as the online algorithms pass over the data.



When deployed on the highly dynamic invivo cardiac perfusion data, our algorithm achieves competitive reconstruction error in less wall clock time. In the tubal-sampling case, which is most practical in real fMRI collection, our method can more rapidly update its subspace estimate during initialization. Beginning at frame 41, strong breathing motion occurs, and the three algorithms are comparable in their subspace tracking abilities. Adjusting the streaming CP algorithms' hyperparameters and STC's choice of multirank also requires exhaustive trial and error, and the results are often sensitive to these choices.

\begin{figure}
\centering
\subfloat[40\% missing entries at frame 39.]{\includegraphics[trim={0cm 0cm 0cm 0cm},clip,width=0.3\textwidth]{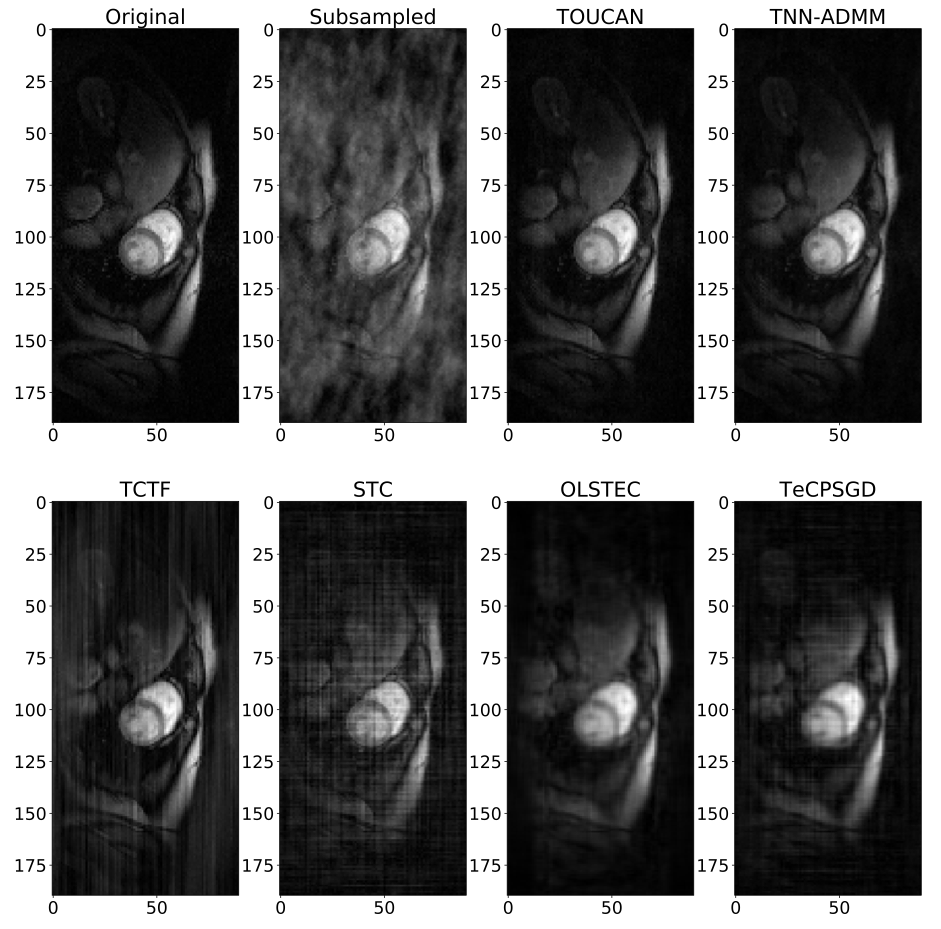}}
\\
\subfloat[ 40\% sampled entries uniformly at random.]{\includegraphics[scale=0.3]{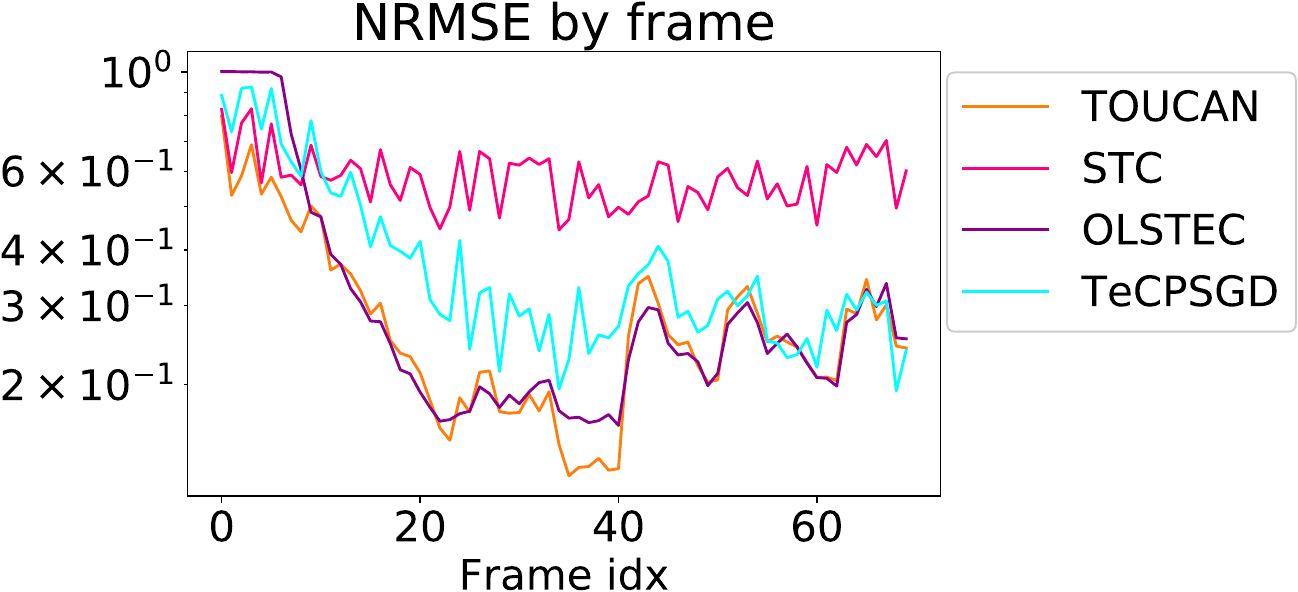}}
\caption{Reconstructed myocardial perfusion images and NRMSE of recovered real component by frame index.}
\label{fig:cardiac_reconstruction_40p}
\end{figure}

\begin{table*}
\centering
\setlength{\tabcolsep}{7pt}
\begin{tabular}{|l|l| *7{c}|}
\hline
{Sample \%}  & {}  & Subsampled/Zero-filled & TOUCAN & TNN-ADMM & TCTF & OLSTEC & TeCPSGD & STC \\
\thickhline
    \multirow{2}{*}{50 - Random} & SSIM & 0.4735 & \it{0.8637} & \bf{0.9518} & 0.7129 & 0.7663 & 0.7709 & 0.6474 \\
    &Time & -- & \it{1.798} & 31.29 & 11.05 & 14.94 & \bf{1.289} & 7.456 \\
    \hline
    \multirow{2}{*}{50 - Tube} & SSIM & 0.5678 & \it{0.8507} & \bf{0.9350} & 0.7118 & 0.7403 & 0.7171 & --\\
    &Time & -- & \bf{0.8240} & 29.22 & 10.51 & 13.75 & \it{1.171} & --\\
\thickhline
    \multirow{2}{*}{40 - Random} & SSIM & 0.3984 & \it{0.8102} & \bf{0.9266} & 0.6384 & 0.7260 & 0.7272 & 0.6412\\
    &Time & -- & \it{2.110} & 32.55 & 11.13 & 14.17 & \bf{1.276} & 6.928 \\
    \hline
    \multirow{2}{*}{40 - Tube} & SSIM & 0.5094 & \it{0.7800} & \bf{0.8984} & 0.6665 & 0.6905 & 0.6547 & --\\
    &Time & -- & \bf{0.7325} & 29.06 & 10.35 & 13.56 & \it{1.046} & --\\
\thickhline
    \multirow{2}{*}{20 - Random} & SSIM & 0.2508 & 0.5623 & \bf{0.8214} & 0.4021 & 0.5609 & \it{0.5719} & 0.5524\\
    &Time & -- & \it{3.132} & 29.88 & 9.997 & 14.29 & \bf{1.116} & 5.130\\
    \hline
    \multirow{2}{*}{20 - Tube} & SSIM & 0.3744 & \it{0.4913} & \bf{0.7315} & 0.4316 & 0.4507 &  0.2817 & --\\
    &Time & -- & \bf{0.6613} & 16.49 & 10.11 & 13.10 & \it{0.9145} & --\\
\thickhline
\end{tabular}
\caption{Invivo myocardial perfusion experiment statistics. Bold indicates best, and italicized indicates second-best.}
\label{tab:mri_stats_cardiac}
\end{table*}

\section{Discussion \& Future Work}
\label{sec:future_work}
In this paper we presented a novel algorithm for low-tubal-rank tensor completion with stochastic gradient descent on the product of Grassmann manifolds under the t-SVD algebraic framework. Our method avoids computing any SVDs, and only needs to update and store a smaller orthonormal tensor and the lateral slice of weights per iteration, leading to a powerful and efficient online algorithm that scales linearly in memory use and computation. TOUCAN naturally extends well-known concepts from matrix algebra to the tensor domain for streaming data under the t-SVD model, making it practical in big data settings where batch methods would become intractable.

\new{As long as the input tubal-rank to our algorithm is an upper bound for the tubal-rank of the data generated, our method should find a good approximation to the tensor, in which case some of the tensor factors may have small coefficients, showing that the rank could be smaller. Establishing good techniques for determining tubal-rank in an online way from missing data is a very interesting direction for future work.}

TOUCAN is practical in many big data problems where the tensor data is inherently oriented, such as time series data, and contains modes with periodic data best captured by the FFT in the t-SVD framework. Fixing the third mode factor matrix to be the DFT matrix is a strong model assumption, and other works have extended the t-SVD to use other fast orthogonal transforms in the third mode \cite{KERNFELD2015545}. Future work could consider learning an orthogonal factor matrix in the third mode that best fits the data.

Choosing the best tensor orientation is not always apparent and requires trial and error. While t-SVD algorithms can leverage periodic structure in the data, CP and Tucker models are compatible with any tensor orientation. These methods may also be preferable when the CP or multilinear ranks are much smaller than either tensor mode dimension; TOUCAN's memory requirement will grow multiplicatively between $d_1$, $d_3$, and $k$ to store the orthonormal basis, whereas CP and Tucker methods require only storing three small factor matrices and a small core tensor in the case of Tucker tensors. Lastly, t-SVD methods are only useful for imputing missing entries when the data reveals a low-tubal-rank structure, but not for recovering interpretable latent factors that may be useful for data analysis. An interesting line of future work would be to develop a novel tensor decomposition agnostic to tensor orientation and enjoys the low memory footprint and latent factor interpretability of CP decompositions.

\section*{Acknowledgments}
We would like to thank the anonymous reviewers for their detailed and thorough feedback that greatly helped improve the precision of our paper. We would like to also acknowledge an anonymous reviewer for pointing out the relationship of the t-SVD to the BTD, and Shuchin Aeron for his helpful advice during the course of this project. This work was supported in part by AFOSR YIP award FA9550-19-1-0026, NSF CAREER award CCF-1845076, NSF BIGDATA award IIS-1838179, and the IAS Charles Simonyi endowment.

\bibliographystyle{siamplain}

\begin{IEEEbiographynophoto}{Kyle Gilman}
Kyle Gilman is a Ph.D. candidate in Electrical and Computer Engineering working with Professor Laura Balzano at the University of Michigan, Ann Arbor, MI. Kyle received his B.S. in Electrical Engineering from the University of Wyoming 2017. His main research focus is on low-rank modeling and optimization for matrix and tensor factorizations, online learning, missing data completion, and applications to signal processing and data science problems. 
\end{IEEEbiographynophoto}
\begin{IEEEbiographynophoto}{Davoud Ataee Tarzanagh}
is currently a Postdoctoral Research Fellow in
the EECS Department at the University of Michigan. He completed his
Ph.D. studies in Mathematics at the University of Florida in 2020. His
current research interests include mathematical optimization, analysis of high
dimensional data with network structure, and tensor data analysis.
\end{IEEEbiographynophoto}
\begin{IEEEbiographynophoto}{Laura Balzano}
Laura Balzano is an associate professor of Electrical Engineering and Computer Science at the University of Michigan. She has a PhD from the University of Wisconsin in ECE. She is currently serving as associate editor of the IEEE Open Journal of Signal Processing and the SIAM Journal of the Mathematics of Data Science. She is recipient of the NSF Career Award, ARO Young Investigator Award, AFOSR Young Investigator Award, and faculty fellowships from Intel and 3M. She received the Vulcans Education Excellence Award at the University of Michigan. Her main research focus is on modeling and optimization with big, messy data — highly incomplete or corrupted data, uncalibrated data, and heterogeneous data — and its applications in a wide range of scientific problems. 
\end{IEEEbiographynophoto}


\newpage
\appendices
\crefalias{section}{appendix}

\section{Proof of Theorem~\ref{thm:sample_bound_coherence}}
\label{proofs:cgd}
\begin{proof}
The proof follows from \cref{lem:avron}, and the fact $\Fkron \Ub$ is a $d_1d_3 \times d_3r$ matrix with orthonormal columns.

From \cref{lem:avron}, we have that 
\begin{equation}
  J \leq \frac{1}{2} \sqrt{\kappa} \log(2 / \epsilon) \leq \frac{1}{2} \left(\frac{1 + \delta^{-1}\tau}{1 - \delta^{-1}\tau}\right)^{\frac{1}{2}} \log(2/\epsilon),   
\end{equation}
where 
\begin{align*}
\kappa :=\kappa(\FOmeg\Ub)^2, ~~\textnormal{and}~\tau  := C \sqrt{d_1d_3 \mu(\Fkron\Ub) \log(|\Omega|)/(|\Omega|)}.
\end{align*}
 \new{
 
 Now, from our assumption bounding the coherence of the iterates $\tU_t$, the result from \cref{eq:coherence_equivalency} gives $\mu(\Fkron\Ub) \leq \mu_0 r / d_1$. To ensure the bound is not vacuous, we must ensure
    \begin{align}\label{eqn:bounn0}
        1 \geq \delta > \tau \geq C\sqrt{d_1 d_3 \mu(\U) \log(|\Omega|) / |\Omega|}.
    \end{align}
 In other words, we must sample sufficiently many rows, i.e., 
 \begin{align}\label{eqn:bounn}
     |\Omega_t| / \log(|\Omega_t|) > C^2 \mu_0 r d_3 .
 \end{align}
 for some constant $C$ independent of $\mu_0$, $r$, $d_1$, and $d_3$.
 
We can simply verify Equation~\eqref{eqn:bounn} for low rank tensors as follows. Without loss of generality, assume $d_1 \geq d_3$. Then, it follows that $\log(|\Omega_t|) \leq 2 \log(d_1) \ll d_1$. Further, if the tensor is low rank and incoherent, both $\mu_0$ and the tubal rank are small, which implies that the right side of Equation~\eqref{eqn:bounn} is sufficiently small. 
Finally \cref{eqn:bounn} ensures that \cref{eqn:bounn0} holds, and we can ensure the number of CGD iterations is low.}
 \end{proof}

\section{Missing Tensor Tubes}
\label{subsec:tube_missing}

Again, let $\tX = [\slice{X}_1 \, \hdots, \slice{X}_{T}] \in \R^{d_1 \times T \times d_3}$ be a set of lateral slices for each time instance. At every time $t$, we observe an incomplete lateral slice $\tX_t \in \mathbb{M}^{d_1}_{{d_3}}$ on the indices $\Omega_t \subset \{1,\hdots,d_1\}$ where not all tubes of the slice are observed. Denote $\tensor{P}_{\Omega_t} \in \R^{|\Omega_t| \times d_1 \times d_3}$ as the tensor that selects the coordinate axes of $\R^{d_1}$ indexed by $\Omega_t$. $\tensor{P}_{\Omega_t}$ is a tensor whose first frontal slice is a subsampled identity matrix on the rows indexed by $\Omega_t$; all other frontal slices are zeros. We then observe the lateral slice $\tensor{P}_{\Omega_t} \tP \slice{X}_t$ at time $t$. Let $\tU_{\Omega_t}$ denote the subtensor of $\tU$ consisting of the tubes indexed by $\Omega_t$, and $\tX_{\Omega_t} = \tensor{P}_{\Omega_t} \tP \slice{X}_t$ denote a lateral slice in $\R^{|\Omega_t| \times 1 \times d_3}$ observed on the tubes indexed by  $\Omega_t$. It can be shown that the objective function can be rewritten as 
\vspace{-2mm}

\begin{equation}
\begin{aligned}
    \min_{[\tU] \in {\mathcal{G}}(r,d_1,d_3)} &\frac{1}{T} \sum_{t=1}^T \newthree{\min_{\slice{W}_t \in \R^{r \times 1 \times d_3}}} \frac{1}{2}\left\|\slice{X}_{\Omega_t} - \tU_{\Omega_t} \tP \slice{W}_t \right\|^2_F.
    \label{eq:toucan_objective_stochastic_tube}
\end{aligned}
\end{equation}


In the Fourier domain, $\overline{\cL}_t$ becomes becomes
\begin{align}
    \label{eq:toucan_objective_missing_tube}
    \overline{\cL}_t(\Ub) = \newthree{\min_{\Wb_t}} \frac{1}{2} \|\Xb_{\Omega_t} - \Ub_{\Omega_t}\Wb_t \|^2_F.
\end{align}

The notation $\Ub_{\Omega_t} \in \C^{|\Omega_t|d_3 \times d_3r}$ denotes the block-diagonal matrix of $\Ub$ consisting of the rows indexed by $\Omega_t$. Similarly, $\Xb_{\Omega_t}$ is a block-diagonal matrix in $\R^{|\Omega_t|d_3 \times d_3}$ observed on the rows indexed by  $\Omega_t$. The problem is block-diagonal, and as the work in \cite{josh_girson} showed, it is separable in each frontal slice in the Fourier domain. The algorithm is similar to that in Alg. \ref{alg:toucan}, except the optimal weights $\slice{W}_t(\tU)$ can be solved exactly in closed form using pseudo-inverses in the Fourier domain, and $\rhob_k$ is replaced by $\rb_k$ in Eq. \ref{eq:Ubar_rank_one_geodesic}. Likewise, our step size is $\eta_k = \arctan(\|\rb_k\|/\|\Wb_{t,k}\|)$. We give the full algorithm in Algorithm \ref{alg:toucan_missing_tubes}.


\begin{algorithm}
\begin{algorithmic}[1]
\REQUIRE{\textbf{Data:} $\slice{X}_t \in \R^{d_1 \times 1 \times d_3} \quad \forall t=1,\hdots,T$ observed on $\Omega_t$; tubal-rank $r$.

\STATE Initialize Fourier transformed orthonormal tensor $\tUb_0 \in \C^{d_1 \times r \times d_3}$.}

\FOR{$t=1$ to $T$}
\STATE Compute $\tXb_{\Omega_t} = \new{\fft(\Delta_{\Omega_t}(\tX_t)},[],3)$.
\STATE Estimate optimal weights: $\Wb_{t,k}(\Ub_t) = {\Ub_{\Omega_t,k}^\dagger} \Xb_{\Omega_t,k}$.

\STATE Predict full vector: $\overline{\PP}_t = \Ub_t \Wb_t(\Ub_t)$.

\STATE Shape into tensor and transform: $\slice{P}_t = \ifft(\overline{\tensor{P}}_t,[],3)$.

\STATE Compute residual: $\slice{R}_t = \Delta_{\Omega_t}(\slice{X}_t) - \slice{P}_t $.

\STATE {Update subspace: } $\tUb_{t+1}$ from \cref{eq:Ubar_rank_one_geodesic}.

\STATE Transform: $\tU_{t+1} = \ifft(\tUb_{t+1},[],3)$.
\STATE Transform: $\slice{W}_t(\tU_t) = \ifft(\tWb_t(\tUb_t),[],3)$.
\ENDFOR

\RETURN{} $\tU, \slice{W}_t(\tU_t), \quad \forall t=1,\hdots,T$

\end{algorithmic}
\caption{Tensor rank-One Update on the Complex grassmanniAN (TOUCAN): Missing Tensor Tubes}
\label{alg:toucan_missing_tubes}
\end{algorithm}

\section{Gradient derivation} \label{appendix:sec:proof_derivative}

Our algorithm substitutes $\wb_t(\Ub)= (\Ub' \FOmeg' \FOmeg \Ub)^{-1}\Ub' \FOmeg'$ into $\overline{\clL}(\Ub)$ and computes the gradient with respect to $\Ub$ directly. First rewrite $\overline{\cL}_t$ as
\begin{align}
    \nonumber 
    \overline{\cL}_t(\Ub) &= \frac{1}{2}\|\FOmeg \xb_t \|_2^2 - \frac{1}{2}\tr((\Ub' \bmC \Ub)^{-1} \Ub' \bmB \Ub ),
      \end{align}
where $\bmC := \FOmeg' \FOmeg$ and $\bmB := \bmC \xb_t \xb_t' \bmC$ for ease of notation.

Now, we can take the gradient with respect to $\Ub$ for this form of trace function via \cite[Equation (126)]{matrix_cookbook} and obtain
\begin{align}
\nonumber     
    \frac{\partial \overline{\cL}_t}{\partial \Ub} &=  -\bmB \Ub (\Ub' \bmC \Ub)^{-1} \\
    &+ \bmC \Ub(\Ub' \bmC \Ub)^{-1} \Ub' \bmB \Ub (\Ub' \bmC \Ub)^{-1}.
\end{align}
It is then straight-forward to see Eq. \cref{eq:grad:U} is equivalent to the above gradient by substituting the expression for $\wb_t(\Ub)$ and simplifying.

\section{t-Product and t-SVD}
\label{appendix:prelims}

Using properties of the Fourier Transform, we give Lemma \cref{lemma:conj_symm}, which describes conjugate symmetry of a real-valued signal transformed into the Fourier domain:

\begin{lemma} \cite{Lu} \label{lemma:conj_symm} Given $\tA \in \R^{d_1 \times d_2 \times d_3}, \Ab_1 \in \R^{d_1 \times d_2} \text{ and } \conj(\Ab_k) = \Ab_{d_3 - k + 2},
    \quad k = 2,\hdots, \lceil \frac{d_3 + 1}{2} \rceil$.
\end{lemma}






\cref{lemma:conj_symm} states the conjugate symmetry property for a real-valued signal in the frequency domain using properties from the Fourier transform; this will be useful later for avoiding redundant computations.

\begin{algorithm}[H]
\hspace*{\algorithmicindent} \textbf{Inputs: }$\tA \in \R^{d_1 \times d_2 \times d_3}, \tB \in \R^{d_2 \times l \times d_3}$ \\
 \hspace*{\algorithmicindent} \textbf{Output: } $\tC = \tA \tP \tB \in \R^{d_1 \times l \times d_3}$ 
 
\begin{algorithmic}[1]
\STATE Compute $\tAb = \fft(\tA,[],3)$ and $\tBb = \fft(\tB,[],3)$
\STATE Compute each frontal slice of $\tCb$ by 
\begin{equation*}
    \Cb_k = 
    \begin{cases}
        \Ab_k\Bb_k, & k = 1,\hdots, \lceil \frac{d_3 + 1}{2} \rceil \\
        \conj(\Cb^{(d_3 - k + 2)}), & k = \lceil \frac{d_3 + 1}{2} \rceil + 1, \hdots, d_3
    \end{cases}
\end{equation*}
\STATE Compute $\tC = \ifft(\tCb,[],3)$

\end{algorithmic}
\caption{Tensor-Tensor Product \cite{Lu}}
\label{alg:t-product}
\end{algorithm}


\begin{definition}{Conjugate transpose} \cite{Kilmer2011FactorizationSF} The conjugate transpose of a tensor $\tA \in \C^{d_1 \times d_2 \times d_3}$ is the tensor $\tA' \in \C^{d_2 \times d_1 \times d_3}$ obtained by conjugate transposing each frontal slice of $\tA$ and then reversing the order of transposed slices 2 through $d_3$:
\begin{equation*}
    \tA' = \fold\left( \begin{bmatrix} {\A_{1}}' & {\A_{d_3}}' & \cdots & {\A_{2}}' \end{bmatrix}'\right).
\end{equation*}
\end{definition}

\begin{definition}{Identity tensor}\cite{Kilmer2011FactorizationSF} The identity tensor $\tensor{I}_{nnd_3} \in \R^{n \times n \times d_3}$ is the tensor whose first frontal slice being the $n \times n$ identity matrix, and all other frontal slices being all zeros. Property: $\tA \tP \tensor{I} = \tensor{I} \tP \tA = \tA$.
\end{definition}

\begin{definition}{Orthogonal tensor} \cite{Kilmer2011FactorizationSF} A tensor $\tensor{Q} \in \R^{n \times n \times d_3}$ is orthogonal if it satisfies $\tensor{Q}' \tP \tensor{Q} = \tensor{Q} \tP \tensor{Q}' = \tensor{I}$.
\end{definition}

\begin{definition}{F-diagonal tensor} \cite{Kilmer2011FactorizationSF} A tensor is called F-diagonal if each of its frontal slices is a diagonal matrix.
\end{definition}
\begin{algorithm}[t]
\hspace*{\algorithmicindent} \textbf{Inputs: }$\tA \in \R^{d_1 \times d_2 \times d_3}$\\
 \hspace*{\algorithmicindent} \textbf{Output: } t-SVD components $\tU,\tensor{S}$, and $\tensor{V}$ of $\tA$. 
 
\begin{algorithmic}[1]
\STATE Compute $\tAb = \fft(\tA,[],3)$
\STATE Compute each frontal slice of $\tUb ,\tSb ,\tVb $ by 
\FOR {$k = 1,\hdots, \lceil \frac{d_3 + 1}{2} \rceil$}
   \STATE $ [\Ub_k, \Sb_k,\Vb_k] = \text{SVD}(\Ab_k)$;
\ENDFOR
\FOR {$k = \lceil \frac{d_3 + 1}{2} \rceil + 1, \hdots, d_3$}
    \STATE $\Ub_k = \conj(\Ub_{d_3 - k + 2}))$
    \STATE $\Sb_k = \conj(\Sb_{d_3 - k + 2}))$
    \STATE $\Vb_k = \conj(\Vb_{d_3 - k + 2}))$
\ENDFOR
\STATE Compute $\tU = \ifft(\tUb ,[],3)$, $\tensor{S} = \ifft(\tSb ,[],3)$, $\tensor{V} = \ifft(\tVb ,[],3)$
\end{algorithmic}
\caption{t-SVD \cite{Lu}}
\label{alg:t-svd}
\end{algorithm}

\newnew{
\section{Proof of Proposition~\ref{prop:well_defined_tensor_grassmannian}}
\label{appendix:proof:prop:well_defined_tensor_grassmannian}
\begin{proof}
{ \color{black}
\ref{itm:prop:one} Assuming we sample enough entries of the data such that $(\FOmeg \Ub)'(\FOmeg \Ub)$ remains full rank, then $\wb_t(\Ub)$ is the unique minimizer of the inner least-squares problem.
\\
\ref{itm:prop:two} For a fixed $\tU$, let $\slice{W}_t(\tU)$ be the unique minimizer in \eqref{eq:bilevel2} as shown in \ref{itm:prop:one},  and say we choose a different basis for $[\tU]$, i.e. $\tU^R := \tU \tP \tR$ for any t-orthogonal tensor $\tR \in \mathcal{O}(r,r,d_3)$. Now we see that $\slice{W}_t(\tU^R) = \tR' \tP \slice{W}_t(\tU)$. Since  $\cL_t(\tU)$ defined in \eqref{eq:bilevel1} merely depends on the product $\tU \tP \slice{W}_t(\tU)$, we have $\clL_t(\tU^R) = \clL_t(\tU)$ which implies that the outer objectives of $\tU$ and $\tU^R$ are identical, i.e., $\clL(\tU^R) = \clL(\tU)$. Hence, the objective is constant over sets of full tubal-rank tensors $\tU$ spanning the same free submodule. Now, considering these sets as an equivalence class $[\tU]$, the problem is well-defined and smooth on the t-Grassmannian. This type of argument was also provided in \cite[Section~3]{boumal2015low} for the offline matrix completion problem on the Grassmannian. 
}

By \cref{propzer:well_defined_tensor_grassmannian}, $\overline{\cL}_t(\Ub)$ in \cref{eq:arb_missing:btd_stochastic_objective2} is a smooth function over the Cartesian product of complex (matrix) Grassmann manifolds in the Fourier domain. This, together with the fact that the Fourier transform operator $\FF_{d_3}$ is invertible, also implies that $\clF :\mathcal{G}(r,d_1,d_3) \rightarrow \mathbb{R}$ is a well-defined smooth function over the product manifold. Further, the solutions of the original problem in \eqref{eq:toucan_objective_incremental2} can be obtained as follows:
\begin{align*}
\tU &= \tUb \times_3 \FF_{d_3}' = \fold(\Ub_1;\Ub_2;\hdots;\Ub_{d_3}) \times_3 \FF_{d_3}',\\ 
\slice{W}_t(\tU) &= \fold(\wb_{t,1};\wb_{t,2};\hdots;\wb_{t,d_3}) \times_3 \FF_{d_3}'.
\end{align*}
This completes the proof.
\end{proof}
}

\section{t-SVD interpretation of TOUCAN}
\label{appendix:sec:tsvd_toucan}
We note here that $\tU$ is one choice of representation for a point on the \new{product} Grassmannian where the Fourier transform along its tubes $\tUb$ has as each frontal face a matrix with orthonormal columns. However, we can equivalently represent this point using an $d_1 d_3 \times d_3r$ block-diagonal matrix in the Fourier domain, with the frontal faces of $\tUb$ on the diagonal. We will revisit this representation below.

We can rewrite the objective function using the block-circulant matrix definition of the t-product:
\begin{align*}
\label{eq:toucan_block_circ_obj}
    \cL_t(\tU)&=  \newthree{\min_{\slice{W}_t}} \frac{1}{2} \|\PP_{\Omega_t} \unfold(\Delta_{\Omega_t}(\slice{X}_t))\\ &\quad \hspace*{2em} \nonumber
    - \PP_{\Omega_t} (\bcirc(\tU) \cdot \unfold(\slice{W}_t)) \|^2_F.
\end{align*}
Here $\PP_{\Omega_t}$ is a subsampled identity matrix of size $|\Omega_t| \times d_1d_3$, $\unfold(\Delta_{\Omega_t}(\slice{X}_t)) \in \R^{d_1d_3}$, $\bcirc(\tU) \in \R^{d_1d_3 \times d_3r}$, and $\unfold(\slice{W}_t) \in \R^{d_3r}$.
Using block-circulant diagonalization and the fact $d_2 = 1$ when processing a single slice, we can rewrite the product $\PP_{\Omega_t} \cdot (\bcirc(\tU) \cdot \unfold(\slice{W}_t))$ as
    \begin{align*}
        &\PP_{\Omega_t} (\FF_{d_3}^{-1} \otimes \I_{d_1})(\FF_{d_3} \otimes \I_{d_1}) \bcirc(\tU) \FF_{d_3}^{-1} \FF_{d_3} \unfold(\slice{W}_t) \\
        &= \PP_{\Omega_t} (\FF_{d_3}^{-1} \otimes \I_{d_1}) \Ub \wb_t,
    \end{align*}
    where $\wb_t := \unfold(\slice{W}_t)$ and
        $\Ub = (\FF_{d_3} \otimes \I_{d_1}) \cdot \bcirc(\tU)\cdot \FF_{d_3}^{-1}$. $\Ub$ is of size $d_1d_3 \times d_3r$ and gives us another representation of $\tUb$, with the frontal slices of $\tUb$ on the diagonal, with $d_3$ blocks of size $d_1 \times r$. We therefore have the following equivalent form for $\overline{\cL}_t(\Ub)$:
    \begin{align*}
\overline{\cL}_t(\Ub) = \newthree{\min_{\wb_t}}\frac{1}{2} \|\FOmeg (\xb_t - \Ub\wb_t)\|^2_2, 
    \end{align*}
    \noindent where $\xb_t \in \R^{d_1d_3} := \texttt{vec}(\Delta_{\Omega_t}(\slice{X}_t) \times_3 \FF_{d_3}) \in \C^{d_3r}$ for convenient notation. Finally, $\FOmeg = \PP_{\Omega_t} (\FF_{d_3}^{-1} \otimes \I_{d_1}) \in \C^{|\Omega_t| \times d_1d_3}$ is the subsampled inverse Fourier transform.




\section{Supporting lemmas of Theorem~\ref{thm:sample_bound_coherence}}

The following lemma utilizes the notion of coherence of an $m \times r$ subspace basis $\U$, defined as $\mu(\U) = \max_{1 \leq i \leq m} \|\mathbf{P}_{\U} \ei\|_2^2$, where $\mathbf{P}_{\U}$ is the orthogonal projection onto $\U$ and $\ei$ is the $i^\text{th}$ standard basis vector \cite{candes_recht}.

\begin{lemma}\cite[Lemma 8.3.3]{avron}
Let $\U$ be an $m \times r$ orthonormal matrix and  $\Ss$ be a random subsampling operator that samples $|\Omega|$ rows from $\U$ uniformly \new{such that 
 $|\Omega|/\log(|\Omega|) \geq C^2 m \mu(\U) $}. Let $C$ be a universal constant, and $\delta \in [0,1]$. Then, with probability at least $1 - \delta$
    \vspace{-2mm}
    \begin{equation}
        \mathbb{E}\{\|\I_r - \frac{m}{|\Omega|} \U'\Ss'\Ss\U \|\} \leq C \sqrt{m \mu(\U) \log(|\Omega|)/|\Omega|} := \tau, \text{ and }  \nonumber
    \end{equation}
     \vspace{-2mm}
    \begin{equation}
        \kappa(\Ss \U) \leq \sqrt{\frac{1 + \delta^{-1}\tau}{1 - \delta^{-1}\tau}}.
    \end{equation}
    \label{lem:avron}
\end{lemma}
\vspace{-4mm}
\new{
\begin{lemma} \label{eq:coherence_equivalency}
Let $\FF_{d_3} = [\bmf_1 \hdots \bmf_{d_3}]$
denote the normalized $d_3 \times d_3$ DFT matrix. Let $\Ub$ be the block-diagonal form of $\tU$ in the Fourier domain. For a tensor $\tU$, define 
\begin{equation}\label{eqn:fudef}
\Fkron \Ub := (\FF_{d_3}^{-1} \otimes \I_{d_1})\Ub.
\end{equation}
Then, we have $\mu(\tU) = \mu(\Fkron \Ub)$, where the function $\mu $ is given in Definition~\ref{defn:cohe}.

\begin{proof}
It follows from Definition~\ref{defn:cohe} and \eqref{eqn:bdiag} that  
    \begin{align}\label{eqn:mu1}
    \nonumber
        \mu(\tU) &= \max_{i=1,\hdots,d_1} \left\| \begin{bmatrix} \Ub_1' & & 0 \\ & \ddots & \\ 0 & & \Ub_{d_3}' \end{bmatrix} \begin{bmatrix} \bme_i \\ \vdots \\ \bme_i \end{bmatrix} \right\|_2^2 \\
        &= \max_{i=1,\hdots,d_1}  \sum_{j=1}^{d_3} \| \Ub'_j\bme_i \|_2^2,
    \end{align}
where $\bme_i$ is the $i^{th}$ standard basis vector in $\R^{d_1}$. Further, from the definition of $\Fkron \Ub$ in \eqref{eqn:fudef}, we have 
\begin{align}\label{eqn:mu2}
        \mu(\Fkron \Ub) = \max_{i=1,\hdots,d_1d_3} \| \Ub' (\FF_{d_3} \otimes \I_{d_1}) \bme_i \|_2^2.
    \end{align}
    
 Denote the $(i,j)$-th entry of the normalized DFT matrix by $f_{ij}$. Through simple algebra, we can see that $(\FF_{d_3} \otimes \I_{d_1}) \bme_i = \bmf_m \otimes \bme_n$ for $i \in \{d_1d_3\}$, $m \in \{d_3\}$, and $n\in \{d_1\}$. This together with \eqref{eqn:mu2} implies that 
    \begin{align*}
    \nonumber 
        \mu(\Fkron \Ub) &= \max_{i=1,\hdots,d_1d_3} \| \Ub' (\FF_{d_3} \otimes \I_{d_1}) \bme_i \|_2^2 \\
            \nonumber 
        &= \max_{\substack{n=1,\hdots,d_1 \\ m=1,\hdots,d_3}} \left \|\begin{bmatrix} \Ub_1' & & 0 \\ & \ddots & \\ 0 & & \Ub_{d_3}' \end{bmatrix} \begin{bmatrix} f_{1m} \bme_n \\ \vdots \\ f_{d_{3}m}\bme_n \end{bmatrix} \right\|_2^2 \\
            \nonumber 
        &= \max_{\substack{n=1,\hdots,d_1 \\ m=1,\hdots,d_3}}  \sum_{j=1}^{d_3} |f_{jm}|^2 \|\Ub_j' \bme_n\|^2_2\\
         &= \max_{n=1,\hdots,d_1}  \sum_{j=1}^{d_3} \|\Ub_j' \bme_n\|^2_2.
    \end{align*}
Here, the second equality uses  \eqref{eqn:mu1} and
the last equality follows from the maximum element of the normalized Fourier transform's vectors being $e^0 = 1$.
\end{proof}
    
\end{lemma}
}

\newnew{
\section{Proof of Theorem~\ref{thm:local_article}} 

\begin{proof}
    The result follows by noting the following steps:
\begin{enumerate}[label=\textnormal{(\roman*)}]
\item \label{proof:s1_article}   Under Assumption~\ref{A2_article}, the tensor problem in \cref{eq:toucan_objective_incremental2} 
which can be re-written as $d_3$ separable optimization problems in the Fourier domain:

\begin{align*}
\min_{[\Ub_{k}] \in \overline{\mathcal{G}}(r,d_1)} \frac{1}{T}\sum_{t=1}^T
\newthree{\min_{\wb_{t,k} \in \C^{r} }} \frac{1}{2}\|\mathcal{P}_{\Omega_t}(\xb_{t,k}) - \mathcal{P}_{\Omega_t}(\Ub_{t,k})\wb_{t,k} \|^2_F.
\end{align*}

\item \label{proof:s2_article}
Under assumption~\ref{A1_article}, each $\slice{X}_t$ is generated as 
  \begin{align*}
      \slice{X}_t = \fold(\Fkron \xb_t), \quad \xb_t = \Ub^* \sbt, \quad \sbt \stackrel{i.i.d}{\sim} \mathcal{C}\mathcal{N}(0,\I_{d_3r}),
  \end{align*}
  where $\Ub^* \in \C^{d_1d_3 \times d_3r}$ is block-diagonal. Hence, we have that each slice $\xb_{t,k} = \Ub_k^* \sbtk$ where $\sbtk \stackrel{i.i.d}{\sim} \mathcal{C}\mathcal{N}(0,\I_{r})$. This follows from the fact that if $\bm{s}_t = \unfold(\slice{S}_t) \sim \mathcal{N}(0,\I_{d_3r})$, then $\sbt$ is just a linear transform of a Gaussian-distributed random variable under the (normalized) Fourier transform matrix.
\item \label{proof:s3_article}
Let $\delta:=0.1/d_3$.  Let $\Ub_{t,k}$ denote the $k^{th}$ block of $\tUb$ at iteration $t$, and let $[\Ub_{t,k}]_{\Omega_t}$ denote the restriction of $\Ub_{t,k}$ to the rows indexed in $\Omega_t$. Since $|\Omega_{t}| \geq q$ for all $t$ and $q$ satisfies \eqref{eqn:samb_article}, for all $k \in [d_3]$, we obtain 
  \begin{align}\label{eqn:samb:kwis_article}
  \nonumber
 q &\geq C_1 \log(d_1)^2 r \mu(\Ub_{k}^*) \log(20rd_3)\\
 &\geq  \frac{C_1}{2} \log(d_1)^2 r \mu(\Ub_{t,k}) \log(20rd_3) 
  \end{align}  
for some $C_1\geq 64/3$, where the last inequality follows from \eqref{eqn:epsi:block_article} and our assumption that $\epsilon_{t,k} \leq  \frac{r}{16d_1}\mu(\Ub^*_k)$; see, \cite[Lemma 2.5]{balzano_wright2015} for more details. Now, it follows from \cite[Lemma 2.8]{balzano_wright2015}, for each $k\in[d_3]$,  
\begin{align}\label{eq:incr:low_article}
\begin{split}
\|[\xb_{t,k}-\pb_{t,k}]_{\Omega_t}\|_2^2  &\geq \frac{|\Omega_t|(1-\xi_{t,k}) - r \mu(\Ub_{t,k})\frac{(1+\beta_{t,k})^2}{1-\gamma_{t,k}}}{d_1}\\ &\hspace{2mm}\cdot\|\xb_{t,k} - \Ub_{t,k} \Ub_{t,k}' \xb_{t,k} \|_2^2,
\end{split}
\end{align}
with probability at least $1-3\delta$. Here, $\pb_{t,k} = \Ub_{t,k} \wb_{t,k}$ where $\wb_{t,k}$ is the optimal weights, 
\begin{align*}
\xi_{t,k} &:=
\sqrt{\frac{2\mu(\vb_{t,k})^2}{|\Omega_t|}  \log\left(\frac{1}{\delta}\right)}, \\
\beta_{t,k} &:= \sqrt{2  \mu(\vb_{t,k}) \log\left(\frac{1}{\delta}\right)},\\
 \gamma_{t,k} &:= \sqrt{\frac{8 r \mu(\Ub_{t,k})}{3|\Omega_t|} \log\left(\frac{2r}{\delta}\right)}.
\end{align*}
Since $\delta=0.1/d_3$,  and $|\Omega_{t}| \geq q$ and $q$ satisfies \eqref{eqn:samb:kwis_article}, we get $ \gamma_{t,k}^2 \leq 1/4$ for all $k \in[d_3]$. This together with \cite[Theorem~2.6]{balzano_wright2015} implies that with probability at least $1-\delta$, 
\begin{equation}\label{eq:bou:eig_article}
\lambda_i ([\Ub_{t,k}]_{\Omega_t}'[\Ub_{t,k}]_{\Omega_t}) \in \left[ 0.5 \frac{|\Omega_t|}{d_1},
1.5\frac{|\Omega_t|}{d_1}\right]    
\end{equation}
for all $ i=1, \ldots, r$, where $\lambda_i$ stands for the $i^{th}$ eigenvalue.

We note that \eqref{equ:mu:coh_article} holds with probability at least $1-\bar{\delta}$. Hence, using the union bound, we have that the bounds~\eqref{equ:mu:coh_article}, \eqref{eq:incr:low_article},~\eqref{eq:bou:eig_article} all hold with probability at least $1-(4\delta+\bar{\delta})=1-(\frac{0.4}{d_3}+\bar{\delta})$. 

For all $k \in[d_3]$, let $\theta_{t,k}$ denote the angle between $\mathcal{R}(\Ub_{t,k})$ and the random observation vector $\xb_{t,k}$, where $\mathcal{R}(\cdot)$ stands for the range. Let $\rb_{t,k}:= \xb_{t,k} - \pb_{t,k}$ denote the residual vector for $k^{th}$ block in the Fourier domain. We note that 
$[\rb_{t,k}]_{\Omega_t^c}=0$. Now, using bounds~\eqref{equ:mu:coh_article}, \eqref{eq:incr:low_article},~\eqref{eq:bou:eig_article}, it follows from~\cite[Lemmas~2.9~and~2.10]{balzano_wright2015} that for each $k\in [d_3]$,
\begin{equation}\label{eq:bou:ang_article}
\frac{\|\rb_{t,k}\|^2}{\|\pb_{t,k}\|^2} \ge (0.32) \frac{q}{d_1} \sin^2 \theta_{t,k}
\end{equation}
with probability at least $1-(\frac{0.4}{d_3}+\bar{\delta})$. 

\item  \label{proof:s4_article} Following~\cite[Section 2.5]{balzano_wright2015}, for all $k\in[d_3]$ we obtain
$
\epsilon_{t+1,k} \le \epsilon_{t,k} -\frac{\|\rb_{t,k}\|^2}{\|\pb_{t,k}\|^2}  + 55 \sqrt{\frac{d_1}{q}} \epsilon_{t,k}^{3/2}.
$
This together with \eqref{eq:bou:ang_article} yields 
\begin{subequations}\label{eqn:high:lin_article}
\begin{align}\epsilon_{t+1,k} & \le \epsilon_{t,k} - 0.32 \frac{q}{d_1} \sin^2 \theta_{t,k}
+ 55 \sqrt{\frac{d_1}{q}} \epsilon_{t,k}^{3/2}\\
&\textnormal{with prob. at least $1-\left(\frac{0.4}{d_3}+\bar{\delta}\right)$}, \nonumber \\
\epsilon_{t+1,k} & \le \epsilon_{t,k}
+ 55 \sqrt{\frac{d_1}{q}} \epsilon_{t,k}^{3/2}~~~~\textnormal{otherwise}. 
\end{align}
\end{subequations}

Let $\pi_t:= \sum_{k=1}^{d_3}\sin^2 \theta_{t,k}$. From Step~\ref{proof:s2_article}, we have $\sbtk \stackrel{i.i.d}{\sim} \mathcal{C}\mathcal{N}(0,\I_{r})$ for each $k \in [d_3]$ which together with \cite[Lemma 2.13]{balzano_wright2015} gives
\begin{align}\label{eqn:exp:pi_article}
 \mathbb{E}[\pi_t]=  \sum_{k=1}^{d_3}  \mathbb{E}[ \sin^2 \theta_{t,k}] =\sum_{k=1}^{d_3} \frac{\epsilon_{t,k}}{r} =\frac{\epsilon_t}{r}.  
\end{align}
where the expectation is with respect to the entries of $\sbtk$ that generate the data, and the last equality follows from \eqref{eq:epsilon_def_article}. 

\item  \label{proof:s5_article} We now put together the theory derived in
Steps~\ref{proof:s1_article}--\ref{proof:s4_article} to demonstrate the expected decrease in $\epsilon_t$ over a single iteration.  Taking a union bound across all $d_3$ blocks in \eqref{eqn:high:lin_article}, we obtain
\begin{align*}
&~~~\epsilon_{t+1} = \sum_{k=1}^{d_3} \epsilon_{t+1,k} \\
&\le
 \sum_{k=1}^{d_3} \epsilon_{t,k} - 0.32 \frac{q}{d_1}  \sum_{k=1}^{d_3} \sin^2 \theta_{t,k} + 55 \sqrt{\frac{d_1}{q}}   \sum_{k=1}^{d_3} \epsilon_{t,k}^{3/2} \\
 &\leq 
 \sum_{k=1}^{d_3} \epsilon_{t,k} - 0.32 \frac{q}{d_1}  \sum_{k=1}^{d_3} \sin^2 \theta_{t,k} + 55 \sqrt{\frac{d_1}{q}}   \left(\sum_{k=1}^{d_3} \epsilon_{t,k}\right)^{3/2}\\
 &= \epsilon_{t} - 0.32 \frac{q}{d_1}  \pi_t + 55 \sqrt{\frac{d_1}{q}}   \epsilon_{t}^{3/2}
\end{align*}
with probability at least $0.6-d_3\bar{\delta}$ while
$\epsilon_{t+1} \le \epsilon_{t}
+ 55 \sqrt{\frac{d_1}{q}} \epsilon_{t}^{3/2}
$ otherwise. Taking the expectation with respect to the randomness of the data and using \eqref{eqn:exp:pi_article}, we obtain
\begin{align*}
\mathbb{E}[\epsilon_{t+1} \, | \, \epsilon_{t}] &\le
\epsilon_{t} - (0.32) (0.6-d_3\bar{\delta}) \frac{q}{d_1r} \epsilon_{t} + 55 \sqrt{\frac{d_1}{q}} \epsilon_{t}^{3/2} \\
 & \leq \left(1 - 0.16(0.6  -d_3\bar{\delta})\frac{q}{d_1r}\right)\epsilon_{t}.
\end{align*}
Here, the last inequality follows since 
\begin{align*}
    55 \sqrt\frac{d_1}{q} \epsilon_t^{1/2}
&\le
55 \sqrt\frac{d_1}{q} (0.0029) (0.6-d_3\bar{\delta}) \frac{q^{3/2}}{d_1^{3/2}r}\\
&\le (0.16)  (0.6-d_3\bar{\delta}) \frac{q}{d_1r},
\end{align*}
where the second inequality uses~\eqref{eqn:epsi:total_article}.
\end{enumerate}
\end{proof}
}

\section{Experiment hyperparameters}
\label{sec:hyperparams}
\new{For the algorithms listed in Section~\ref{sec:experiments}, we used the following grids of values for the parameter search:
\begin{itemize}
    \item  \textbf{TeCPSGD grids}: $\text{rank}=[1,5,10,15,20,25,30,50]$, $\lambda = [10^{-5},10^{-4},10^{-3},10^{-2},0.1,0.5,1]$, step size = $[1,10,10^2,10^3,10^4,10^5,10^6]$.

\item \textbf{OLSTEC grids}: $\text{rank}=[1,5,10,15,20,25,30,50]$, $\lambda = [10^{-5},10^{-4},10^{-3},10^{-2},0.1,0.5,1]$, $\mu = [10^{-8},10^{-5},10^{-4},10^{-3},10^{-2},0.1,1]$.

\item \textbf{STC grid}: multi-ranks $(r_1,r_2,r_3)$ where each $r_i$ ranges from $[1,5,10,15,20,25,30]$.

\item \textbf{TCTF grid}: tubal rank ranging from $[1,5,7,10,15,20,25,30]$.
\end{itemize}
}










\end{document}


\title{Supplement to Grassmannian Optimization for Online Tensor Completion and Tracking with the t-SVD}

\author{%
  Kyle~Gilman,~\IEEEmembership{Student~Member,~IEEE,}\\
  and~Laura~Balzano,~\IEEEmembership{Senior~Member,~IEEE}
}
\onecolumn{
\maketitle

\section{Empirical results for CGD iteration bound}

We demonstrate that TOUCAN empirically obeys our derived upper bound from Theorem IV. We generate small problems of synthetic data of size $d_1 = 50, d_2 = 500, d_3 = 20$ with tubal-ranks $k=2$ and $k=3$ from a synthetic FSM $\tU \in \R^{d_1 \times k \times d_3}$, $\tU^* \tP \tU = \tensor{I}_{kkd_3}$. We construct the full matrix $\FOmeg \Ub$ and compute the average $\kappa(\FOmeg \Ub)^2$ for 20 random sampling patterns for each sampling rate and use this to compute the CGD sample bound. We also plot our bound derived in Theorem IV. Similarly, we generate 20 random sampling patterns and TOUCAN processes over each undersampled data set incrementally in two passes, starting from the same random initialization of $\tU_0$ each trial. We record the average number of CGD iterations per algorithm iteration over the 20 trials and plot the results in Fig. \ref{fig:cgd_iter_plot} on the vertical axis of the left figure and the recovered tensor NRMSE on the vertical axis of the right figure; the horizontal axis is ${dof} / |\Omega|$ where $dof = d_3k((d_1 + d_2) - k)$ is the number of degrees of freedom of the set of tubal-rank $k$ tensors and $|\Omega|$ is the number of observed samples.

We observe two regions of interest-- one where the number of CGD iterations is near-linear as a function of the sampling rate, and one where the number of iterations exponentially increases as the data becomes highly undersampled. It is in this transition we see TOUCAN's completion ability begins to fail. While our bound is fairly tight in the linear region, it is quite loose in the highly-undersampled regime. We note that TOUCAN also learns $\tU$, so our assumption that the coherence of $\tU_t$ remains bounded at each iteration appears reasonable and well-evidenced.

\begin{figure}
    \centering
    \subfloat[$k=2, d_1=50,d_2 = 500,d_3=20$]{\includegraphics[trim={3cm 0 0cm 0},clip,scale=0.4]{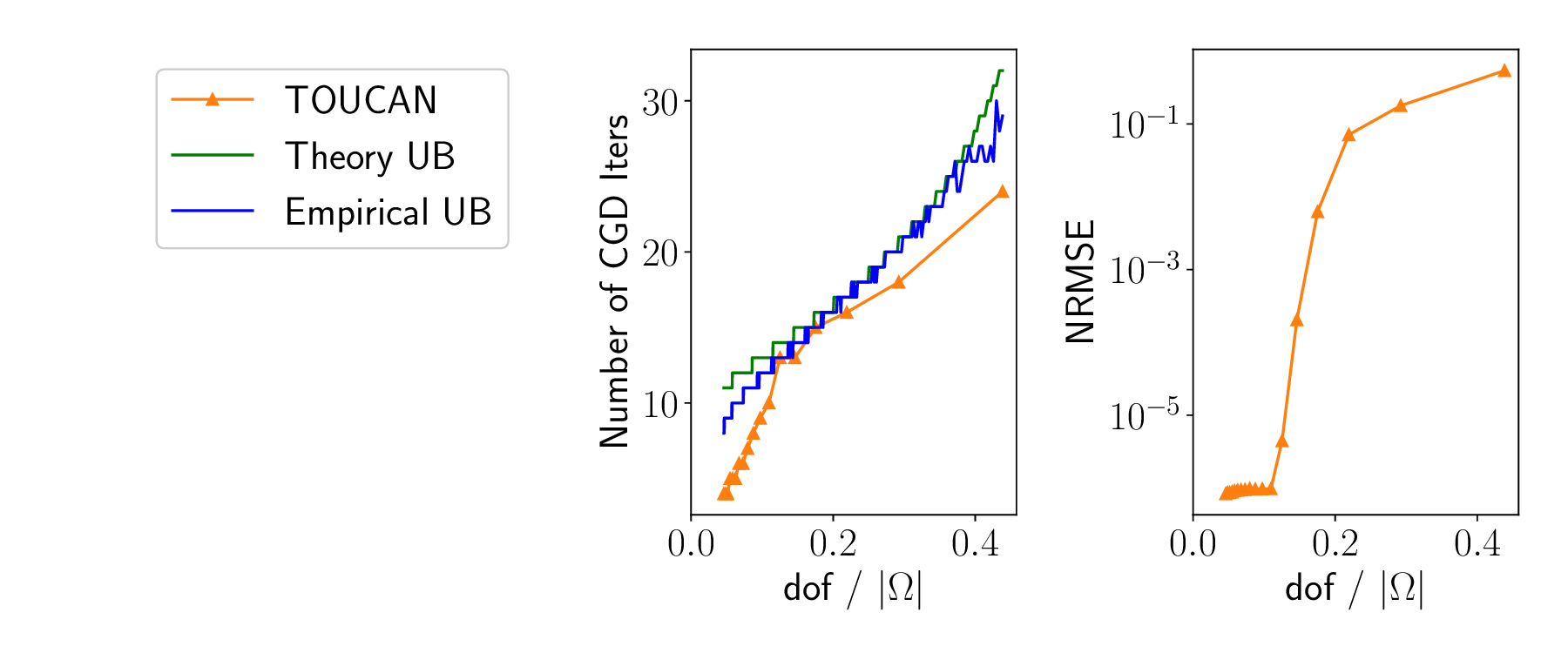}}
    \\
    \subfloat[$k=3, d_1=50,d_2 = 500,d_3=20$]{\includegraphics[trim={10cm 0 0 0},clip,scale=0.4]{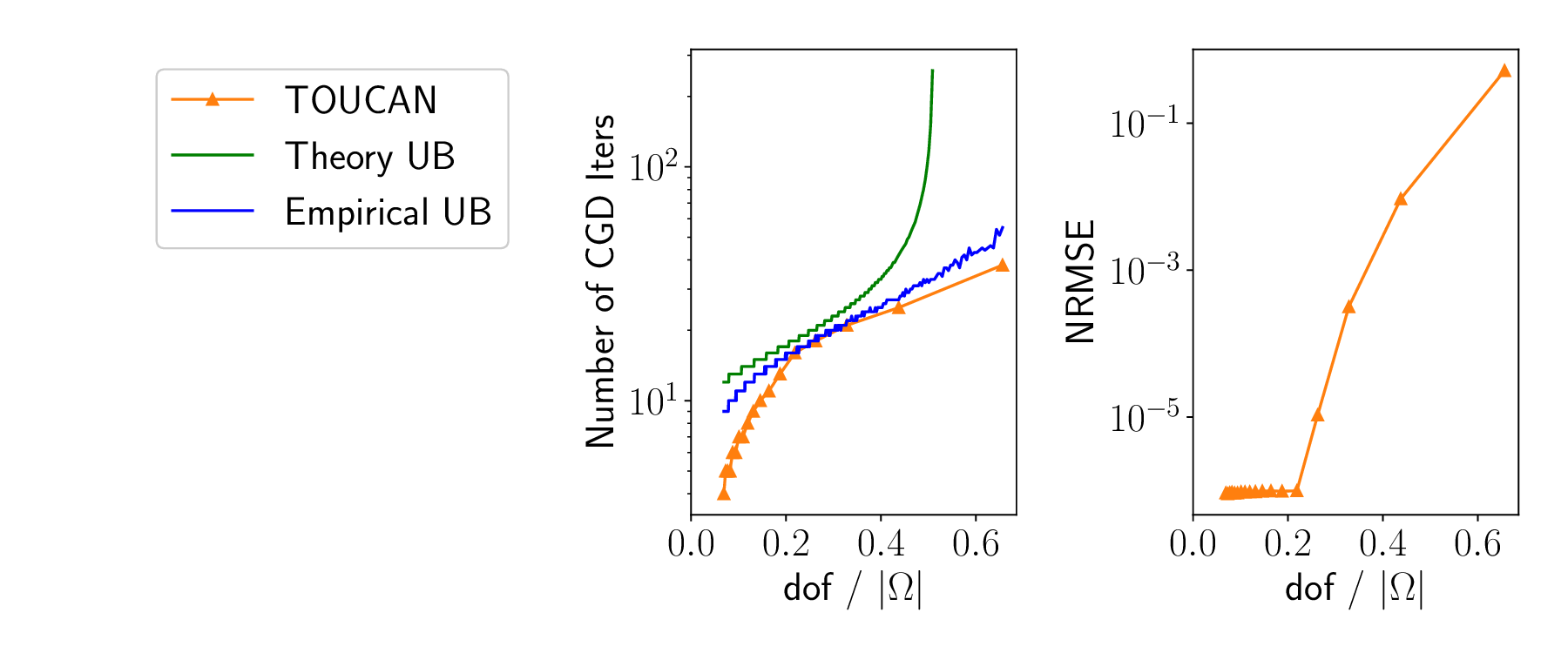}}
    \caption{For our derived bound, we use $\epsilon=10^{-6}, \delta = 0.05$, and $C = 0.00105 \sqrt{d_1d_3}$. On the left, we show the number of conjugate gradient iterations predicted by our upper bound and by the empirically computed condition number, as well as the observed empirical number of CGD iterations by our algorithm. On the right, we show the NRMSE $\|\hat \tX - \tX\|_F / \|\tX\|_F$ of the recovered tensor.}
    \label{fig:cgd_iter_plot}
\end{figure}
  }